\documentclass[11pt, letterpaper]{article}
\usepackage[margin=1in]{geometry}

\usepackage{amsfonts, amsmath, amssymb, amsthm}
\usepackage{cite}
\usepackage{color}
\usepackage{euscript}
\usepackage{graphicx}
\usepackage{mathrsfs} 
\usepackage{microtype}
\usepackage[normalem]{ulem}
\usepackage{wrapfig}

\theoremstyle{definition}
\newtheorem{example}{Example}[section]

\def\ttt{\texttt}

\def\extraspacing{\vspace{3mm} \noindent}
\def\figcapup{\vspace{-1mm}}
\def\figcapdown{\vspace{-0mm}}

\def\vgap{\vspace{1mm}}

\def\tabpos{\hspace{4mm} \= \hspace{4mm} \= \hspace{4mm} \= \hspace{4mm} \=
\hspace{4mm} \= \hspace{4mm} \= \hspace{4mm} \= \hspace{4mm} \= \hspace{4mm}
\kill}
\newcommand{\mytab}[1]{\begin{tabbing}\tabpos #1\end{tabbing}}

\newtheorem{theorem}{Theorem}
\newtheorem{lemma}[theorem]{Lemma}

\newcommand{\boxminipg}[2]{\begin{center}\fbox{\begin{minipage}{#1}#2\end{minipage}}\end{center}}

\newcommand{\myitems}[1]{\begin{itemize}\setlength #1 \end{itemize}}
\newcommand{\myenums}[1]{\begin{enumerate}\setlength #1 \end{enumerate}}

\newcommand{\myfigg}[2]{\begin{figure}\centering #1 \figcapup \caption{#2} \figcapdown \end{figure}}

\newcommand{\bm}[1]{\textrm{\boldmath${#1}$}}

\newcommand{\myeqn}[1]{\begin{eqnarray}#1\end{eqnarray}}

\newcommand{\set}[1]{\{#1\}}

\def\eps{\epsilon}
\def\fr{\frac}
\def\-{\mbox{-}}

\def\nn{\nonumber}

\def\bigmid{\textrm{ $\Big|$ }}

\def\*{\star}

\def\done{\qed \vspace{2mm}}	

\def\figcapup{\vspace{-2mm}}
\def\figcapdown{\vspace{-3mm}}
\def\extraspacing{\vspace{4mm} \noindent}
\def\vgap{\vspace{2mm}}

\def\att{\textrm{\bf att}}
\def\attset{\mathit{attset}}
\def\big{\textrm{big}}
\def\bfr{\textrm{before}}
\def\dom{\textrm{\bf dom}}

\def\join{\mathit{Join}}
\def\map{\mathit{map}}
\def\non{\mathit{non}}

\def\root{\mathit{root}}
\def\scheme{\mathit{scheme}}
\def\sigp{\textrm{sigpath}}
\def\sml{\textrm{small}}
\def\super{\mathit{super}}
\def\tmp{\textrm{tmp}}

\def\T{\mathcal{T}}

\title{Parallel Acyclic Joins with Canonical Edge Covers}

\author{Yufei Tao\\
Department of Computer Science and Engineering \\
Chinese University of Hong Kong \\
taoyf@cse.cuhk.edu.hk}

\begin{document}
\begin{sloppy}
\maketitle
\begin{abstract}
    In PODS'21, Hu presented an algorithm in the massively parallel computation (MPC) model that processes any acyclic join with an asymptotically optimal load. In this paper, we present an alternative analysis of her algorithm. The novelty of our analysis is in the revelation of a new mathematical structure --- which we name {\em canonical edge cover} --- for acyclic hypergraphs. We prove non-trivial properties for canonical edge covers that offer us a graph-theoretic perspective about why Hu's algorithm works. 
\end{abstract} 

\vspace{30mm}
\noindent {\bf Keywords:} Joins, Conjunctive Queries, MPC Algorithms, Parallel Computing. 

\vspace{5mm}

\noindent {\bf Acknowledgments:} This work was partially supported by GRF projects 142078/20 and 142034/21 from HKRGC. 

\pagebreak 

\section{Introduction} \label{sec&intro} 

Massively parallel join processing has attracted considerable attention in recent years. This line of research makes two types of contributions. The first consists of algorithms that promise excellent performance. The second, more subtle, type of contributions comprises knowledge revealing {\em mathematical structures} in the underlying problems. The latter is a necessary side-product of the former. In general, as human beings switch to a more generic setting, their knowledge from restrictive settings often proves insufficient, which then necessitates deeper investigation into the problem characteristics. Traditional studies  have focused on joins in the RAM computation model \cite{abs21,knrr15,nnrr14,nprr18,nrr13}, a degenerated ``parallel'' setup having only one machine. Designing algorithms to work with any number of machines poses serious challenges and demands novel findings \cite{au11,bks17,h21,hyt19,kbs16,ks17,t20,kst20,qt21} beyond the RAM literature. 

\vgap 

This paper will focus on {\em acyclic joins}, a class of joins with profound importance in database systems \cite{ahv95,bks17,hy19,h21,iuv17,y81}. Recently, Hu \cite{h21} developed a worst-case optimal massively parallel algorithm for acyclic joins. In the current work, we will provide an alternative, hopefully more accessible, analysis of her elegant algorithm. The real excitement from our analysis is the identification of a new mathematical structure --- we call ``canonical edge cover'' --- for acyclic hypergraphs. The structure reveals a unique characteristic of acyclic joins and is a core reason why Hu's algorithm works. 


\subsection{Problem Definition} \label{sec&intro&prob}

\noindent {\bf Acyclic Joins.} Let $\att$ be a set where each element is called an {\em attribute}. Let $\dom$ be another set where each element is called a {\em value}. We assume a total order on $\dom$; if not, manually impose one by ordering the values arbitrarily. 

\vgap

A {\em tuple} over a set $U \subseteq \att$ is a function $\bm{u}: U \rightarrow \dom$. For each attribute $X \in U$, we refer to $\bm{u}(X)$ as the {\em value} of $\bm{u}$ on $X$. Given a subset $U' \subseteq U$, define $\bm{u}[U']$ --- the {\em projection} of $\bm{u}$ on $U'$ --- as the tuple $\bm{u'}$ over $U'$ such that $\bm{u'}(X) = \bm{u}(X)$ for every $X \in U'$. A {\em relation} is a set $R$ of tuples over the same set $U$ of attributes. We call $U$ the {\em scheme} of $R$, a fact denoted as $\scheme(R) = U$. If $U$ is the empty set $\emptyset$, then $R$ is also $\emptyset$. 

\vgap 

We represent a {\em join query} (henceforth, simply a ``join'' or ``query'') as a set $Q$ of relations. Define $\attset(Q) = \bigcup_{R \in Q} \scheme(R)$. The query result --- denoted as $\join(Q)$ --- is the following relation over $\attset(Q)$
\myeqn{
    \join(Q) = \left\{ 
    \textrm{tuple $\bm{u}$ over $\attset(Q)$} \bigmid \forall R \in Q, \, \textrm{$\bm{u}[\scheme(R)] \in R$} 
    \right\}. 
    \nn 
}
If the relations in $Q$ are $R_1, R_2, ..., R_{|Q|}$, we may represent $\join(Q)$ also as $R_1 \bowtie R_2 \bowtie ... \bowtie R_{|Q|}$.

\vgap

$Q$ can be characterized by a hypergraph $G = (V, E)$ where each vertex in $V$ is a distinct attribute in $\attset(Q)$, and each hyperedge in $E$ is the scheme of a distinct relation in $Q$. $E$ may contain identical hyperedges because two (or more) relations in $Q$ can have the same scheme. The term ``hyper'' suggests that a hyperedge can have more than two attributes. 

\vgap 

\myfigg{
    \includegraphics[height=40mm]{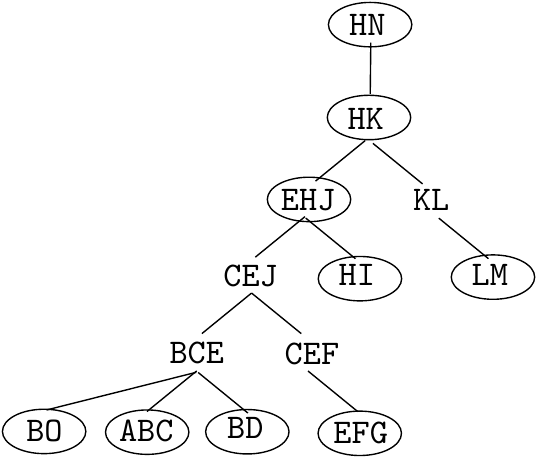}
}{A hyperedge tree example\label{fig&intro-ex}}


A query is {\em acyclic} if its hypergraph is acyclic. Specifically, a hypergraph $G = (V, E)$ is {\em acyclic} if we can create a tree $T$ where
\myitems{
    \item every node in $T$ stores (and, hence, ``corresponds to'') a distinct hyperedge in $E$; 
    \item {{\bf (connectedness requirement)}} for every attribute $X \in V$, the set $S$ of nodes whose corresponding hyperedges contain $X$ forms a connected subtree in $T$. 
}
We will call $T$ a {\em hyperedge tree} of $G$ (also known as the {\em join tree} of $Q$ in the literature).

\begin{example} \label{ex&intro}
    Consider the hypergraph $G = (V, E)$ where $V = \set{\ttt{A},\ttt{B},...,\ttt{O}}$ and $E =$ $\{\ttt{ABC}, \ttt{BD},$ $\ttt{BO},$ $\ttt{EFG},$ $\ttt{BCE},$ $\ttt{CEF}, \ttt{CEJ}, \ttt{HI}, \ttt{LM}, \ttt{EHJ}, \ttt{KL}, \ttt{HK}, \ttt{HN}\}$. Figure~\ref{fig&intro-ex} shows a hyperedge tree $T$ of $G$. To understand the connectedness requirement, observe the connected subtree formed by the five hyperedges involving \ttt{E}. \done
\end{example}

As $G$ and $T$ both contain ``vertices'' and ``edges'', for better clarity we will obey several conventions throughout the paper. A vertex in $G$ will always be referred to as an {\em attribute}, while the term {\em node} is reserved for the vertices in $T$. Furthermore, to avoid confusion with hyperedges, we will always refer to an edge in $T$ as a {\em link}. 

\vgap

We use $m$ to denote the {\em input size} of $Q$, defined as $\sum_{R \in Q} |R|$, namely, the total number of tuples in the relations participating in the join. 

\extraspacing {\bf Computation Model.} We assume the {\em massively parallel computation} (MPC) model which is popular in designing massively parallel algorithms \cite{au11,bks17,h21,hyt19,kbs16,ks17,t20,kst20,qt21}. In this model, we have $p$ machines, each storing $\Theta(m/p)$ tuples from the relations of a query $Q$ initially. An algorithm executes in {\em rounds}, each having two phases: in the first phase, each machine performs local computation; in the second, the machines exchange messages (every message must have been generated at the end of the first phase). An algorithm must finish in a constant number of rounds, and when it does, every tuple in $\join(Q)$ must reside on at least one machine. The {\em load of a round} is the largest number of words received by a machine in that round. The {\em load of an algorithm} is the maximum load of all the rounds. The objective is to design an algorithm with the smallest load. 

\extraspacing {\bf Math Conventions.} The number $p$ of machines is considered to be at most $m^{1-\eps}$, for some arbitrarily small constant $\eps > 0$. Every value in $\dom$ can be represented with $O(1)$ words. Our discussion focuses on {\em data complexities}, namely, we are interested in the influence of $m$ on algorithm performance. For that reason, we assume that the hypergraph $G$ of $Q$ has $O(1)$ vertices. Given an integer $x \ge 1$, the notation $[x]$ represents the set $\set{1, 2, ..., x}$. 

\subsection{Previous Results} \label{sec&intro&prev}


\noindent {\bf Fractional Edge Coverings and the AGM bound.} Consider a query $Q$ (which may or may not be acyclic) with hypergraph $G = (V, E)$. Associate every hyperedge $e \in E$ with a real-valued {\em weight} $w_e$, which falls between 0 and 1. Impose a constraint on every attribute $X \in V$: $\sum_{e \in E: X \in e} w_e \ge 1$, i.e., the total weight of all the hyperedges covering $X$ must be at least 1. A set of weights $\{w_e \mid e \in E\}$ fulfilling all the constraints is a {\em fractional edge covering} of $G$. If we define $\sum_{e \in E} w_e$ as the {\em total weight} of the fractional edge covering, the {\em fractional edge covering number} of $G$ --- denoted as $\rho$ --- is the minimum total weight of all possible fractional edge coverings. A fractional edge covering is {\em optimal} if its total weight equals $\rho$. 


\vgap

The {\em AGM bound}, proved by Atserias, Grohe, and Marx \cite{agm13}, states that the size of $\join(Q)$ is always bounded by $O(m^\rho)$; recall that $m$ is the input size of $Q$. Furthermore, the bound is tight: in the worst case, $|\join(Q)|$ can indeed reach $\Omega(m^\rho)$ \cite{agm13}. 

\extraspacing {\bf Simplification for Acyclic Queries: Edge Covers.} When $Q$ is acyclic, $G = (V, E)$ always admits an optimal fractional edge covering with {\em integral} weights  \cite{h21}. Recall that all the weights $w_e$ ($e \in E$) must fall between 0 and 1. Hence,  every weight in an optimal fractional edge covering must be either 0 or 1. This pleasant property allows the reader to connect $\rho$ to edge ``covers''. A subset $S \subseteq E$ is an {\em edge cover}\footnote{In case the reader is wondering, the literature uses the words ``covering'' and ``cover'' exactly the way they are used in our paper.} of $G$ if every attribute of $V$ appears in at least one hyperedge of $S$. Thus, the value of $\rho$ is simply the minimum size of all edge covers, namely, the smallest number of hyperedges that we must pick to cover all the attributes. 

\extraspacing {\bf Join Algorithms in RAM.} An algorithm able to answer $Q$ using $O(m^\rho)$ time in the RAM model is worst-case optimal. Indeed, as $|\join(Q)|$ can be $\Omega(m^\rho)$, we need $\Theta(m^\rho)$ time just to output $\join(Q)$ in the worst case. Ngo et al.\ \cite{nprr12} designed the first algorithm that guarantees a running time of $O(m^\rho)$ for all queries. Since then, the community has discovered more algorithms \cite{abs21,knrr15,nnrr14,nprr18,nrr13} that are all worst-cast optimal (sometimes up to a polylogarithmic factor) but differ in their own features. For an acyclic $Q$, an algorithm due to Yannakakis \cite{y81} achieves a stronger sense of optimality: his algorithm runs in $O(m + |\join(Q)|)$ time, which is clearly the best regardless of $|\join(Q)|$. 

\extraspacing {\bf Join Algorithms in MPC.} Koutris, Beame, and Suciu \cite{kbs16} showed that, in the MPC model, the AGM bound implies a worst-case lower bound of $\Omega(m/p^{1/\rho})$ on the load of any algorithm that answers a query $Q$, where $m$ is the input size of $Q$ and $\rho$ is the fractional edge covering number of the hypergraph $G = (V, E)$ defined by $Q$. 

\vgap 

The above negative result has motivated considerable research looking for MPC algorithms whose loads are bounded by $O(m/p^{1/\rho})$, ignoring polylogarithmic factors; such algorithms are worst-case optimal. The goal has been realized only on four query classes. The first consists of all the cartesian-product queries (i.e., the relations in $Q$ have disjoint schemes); see \cite{au11,bks17c,kst20} for several optimal algorithms on such queries. The second is the so-called Loomis-Whitney join, where $E$ consists of all the $|V|$ possible hyperedges of $|V|-1$ attributes; see \cite{kbs16} for an optimal algorithm for such joins. The third class includes every query where all the hyperedges in $G$ contain {\em at most} two attributes; see \cite{ks17,t20,kst20} for the optimal algorithms. The fourth class comprises all the acyclic queries, which were recently solved by Hu \cite{h21} optimally. It is worth pointing out that Hu's algorithm subsumes an earlier algorithm of \cite{hy19} which is worst-case optimal on a subclass of acyclic queries.

\vgap

Although it still remains elusive what other query classes can be settled with load $O(m/p^{1/\rho})$, now we know that this is {\em unachievable} for certain queries. In \cite{h21}, Hu constructed a class of queries for which every algorithm must incur a load of $\omega(m/p^{1/\rho})$ in the worst case. The result of \cite{h21} suggests that additional parameters --- other than $m, p$, and $\rho$ --- are needed to describe the worst-case optimality of an ideal MPC algorithm. We will not delve into the issue further because it does not apply to acyclic queries (the focus of this paper), but the reader may consult the recent works \cite{h21,qt21} for the latest development on that issue. Finally, we remark that several algorithms \cite{ajr+17,hy19,hyt19} are able to achieve a load sensitive to the join size $|\join(Q)|$. 

\subsection{Our Contributions} \label{sec&intro&ours}

The first, easy-to-discern, contribution of our paper is a new analysis of Hu's algorithm \cite{h21} for acyclic queries. Our second contribution is the introduction of {\em canonical edge cover} as a mathematical structure inherent in acyclic queries. We prove a suite of graph-theoretic properties for canonical edge covers and use them to give a more fundamental interpretation of the design choices in Hu's algorithm. The rest of the section will provide an overview of our results and techniques. 

\extraspacing {\bf Clustering, $\bm{k}$-Groups, $\bm{k}$-Products, and Induced Loads.} We first create a conceptual framework to state Hu's and our results on a common ground. Define a {\em clustering} of $E$ (the hyperedge set of $G$) as a set $\{E_1, E_2, ..., E_s\}$ for some $s \ge 1$ where (i) each $E_i$ is a subset of $E$, $i \in [s]$, and (ii) $\bigcup_i E_i = E$. We call each $E_i$ a {\em cluster}; note that the clusters need {\em not} be disjoint. 

\vgap

Fix an arbitrary clustering $C = \{E_1, E_2, ..., E_s\}$. Given an integer $k \ge 1$, define a {\em $k$-group} of $C$ as a collection of $k$ hyperedges, each taken from a distinct cluster. 

\begin{example} \label{ex&intro-C} 
    Let $G = (V, E)$ be the hypergraph in Example~\ref{ex&intro} (Figure~\ref{fig&intro-ex}). $C =$ $\{\{\ttt{BO},$ $\ttt{BCE},$ $\ttt{CEJ}\},$ $\{\ttt{ABC},$ $\ttt{BCE},$ $\ttt{CEJ}\},$ $\{\ttt{BD},$ $\ttt{BCE},$ $\ttt{CEJ}\},$ $\{\ttt{EFG},$ $\ttt{CEF}$, $\ttt{CEJ}\},$ $\{\ttt{HI}\}$, $\{\ttt{EHJ}\}$, $\{\ttt{LM},$ $\ttt{KL}\},$ $\{\ttt{HK}\},$ $\{\ttt{HN}\}\}$ is a clustering of $E$. A 3-group example is $\set{\ttt{ABC}, \ttt{BD}, \ttt{EFG}}$. Note that the hyperedges in a $k$-group need not be distinct. For example, $\set{\ttt{CEJ}, \ttt{CEJ}, \ttt{CEJ}}$ is also a 3-group: the first \ttt{CEJ} is taken from the cluster $\set{\ttt{ABC}, \ttt{BCE}, \ttt{CEJ}}$, the second from $\set{\ttt{BD}, \ttt{BCE}, \ttt{CEJ}}$, and the third from $\set{\ttt{EFG}, \ttt{CEF}, \ttt{CEJ}}$. For a non-example, $\set{\ttt{ABC}, \ttt{LM}, \ttt{KL}}$ is not a 3-group. \done
\end{example}

For each hyperedge $e \in E$, let $R(e)$ represent the relation in $Q$ whose scheme is $e$. Given a $k$-group $K$ of the clustering $C$, we define the {\em $Q$-product} of $K$ as $\prod_{e \in K} |R(e)|$ (i.e., the cartesian-product size of all the relevant relations). Given an integer $k$, we define the {\em max $(k,Q)$-product} of $C$ --- denoted as $P_k(Q,C)$ --- as the maximum $Q$-product of all the $k$-groups of $C$. 

\begin{example}
    Continuing on the previous example, the $Q$-product of the 3-group $\set{\ttt{ABC}, \ttt{BCE}, \ttt{CEJ}}$ is $|R(\ttt{ABC})| \cdot |R(\ttt{BCE})| \cdot |R(\ttt{CEJ})|$, while that of the 3-group $\set{\ttt{CEJ}, \ttt{CEJ}, \ttt{CEJ}}$ is $|R(\ttt{CEJ})|^3$. \done
\end{example}

Define the {\em $Q$-induced load} of $C$ as 
\myeqn{
    \max_{k=1}^s \left(P_k(Q,C) / p\right)^{1/k}
    \label{eqn&clustering-load}
}
As $P_k(Q,C) \le m^k$ for any $k \in [s]$, it must hold that $\eqref{eqn&clustering-load} \le m / p^{1/s}$. 

\vgap

We can now give a more detailed account of Hu's result \cite{h21}. She proved that the load of her algorithm is bounded by $O(L)$, where $L$ is the $Q$-induced load of a {\em certain} clustering with size $s = \rho$, and $\rho$ is the fractional edge covering number of $G$. It thus follows immediately that $L \le m/p^{1/s}$. In \cite{h21}, Hu presented a recursive procedure to identify the clustering $C$ whose $Q$-induced load equals the target $L$. The procedure, however, is somewhat sophisticated, making it difficult to describe the target $C$ in a succinct manner. Such difficulty is unjustified, especially given the algorithm's elegance, and indicates the existence of a hidden mathematical structure. 

\extraspacing {\bf Our Results and Techniques.} A hypergraph $G$ can have many optimal edge covers (all of which must have size $\rho$). While Hu's analysis \cite{h21} assumes an arbitrary optimal edge cover, we will be choosy about what we work with. In Figure~\ref{fig&intro-ex}, the 9 circled nodes constitute a {\em canonical edge cover} $F$ of $G$. Let us give an informal but intuitive explanation of how to construct this $F$. After rooting the tree in Figure~\ref{fig&intro-ex} at \ttt{HN}, we add to $F$ all the leaf nodes: \ttt{BO}, \ttt{ABC}, \ttt{BD}, \ttt{EFG}, \ttt{HI}, \ttt{LM}. Then, we process the non-leaf nodes bottom up. In processing \ttt{BCE}, we ask: which attributes will disappear as we ascend further in the tree? The answer is \ttt{B}, which is thus a ``disappearing'' attribute of \ttt{BCE}. Then, we ask: does $F$ already cover \ttt{B}? The answer is yes, due to the existence of \ttt{BO}; we therefore do {\em not} include \ttt{BCE} in $F$. We process \ttt{BCE}, \ttt{CEF}, and \ttt{CEJ} similarly, none of which enters $F$. At \ttt{EHJ}, we find  disappearing attributes \ttt{E} and \ttt{J}. In general, as long as one disappearing attribute has not been covered by $F$, we pick the node; this is why \ttt{EHJ} is in $F$. The other nodes \ttt{HK} and \ttt{HN} in $F$ are chosen based on the same reasoning. 

\vgap 

We show that a canonical edge cover determined this way has appealing properties which fit the recursive strategy behind Hu's algorithm very well. At a high level, Hu's algorithm works by simplifying $G$ into a number of ``residual'' hypergraphs to be processed recursively. Interestingly, with trivial modifications (such as removing the attributes that have become irrelevant), a canonical edge cover of $G$ {\em remains canonical on every residual hypergraph}. This is the most crucial property we utilize to relate the load of the original query to those of the ``residual queries'' in forming up a working recurrence. 

\vgap

Our techniques also provide a simple and natural way to pinpoint a clustering $C$ that can be used to bound the algorithm's load. Consider the canonical edge cover $F$ shown in Figure~\ref{fig&intro-ex} (the circled nodes). For each node in $F$, take a ``signature path'' by walking up and stopping right before reaching its lowest proper ancestor in $F$. For example, the signature path of \ttt{ABC} is \{\ttt{ABC}, \ttt{BCE}, \ttt{CEJ}\} (note: the path does not contain \ttt{EHJ}). Likewise, the signature path of \ttt{LM} is \{\ttt{LM}, \ttt{KL}\}. The signature paths of all the nodes in $F$ together produce the clustering $C$ given in Example~\ref{ex&intro-C}. Our main result (Theorem~\ref{thm&main}) states that the $Q$-induced load of $C$ is an upper bound on the load of Hu's algorithm. Because $C$ has a size at most $\rho$, the algorithm's load is thus bounded by $O(m/p^{1/\rho})$.

\section{Canonical Edge Covers for Acyclic Hypergraphs} \label{sec&cover} 

This section is purely graph theoretic: we will establish several new properties for acyclic hypergraphs. Let $G = (V, E)$ be an acyclic hypergraph. A hyperedge $e_1 \in E$ is {\em subsumed} if it is a subset of another hyperedge $e_2 \in E$, i.e., $e_1 \subseteq e_2$.  If an attribute $X$ appears in only a single hyperedge, we call $X$ an {\em exclusive attribute}; otherwise, $X$ is {\em non-exclusive}. Unless otherwise stated, we allow $G$ to be an arbitrary acyclic hypergraph. In particular, this means that $E$ can contain two or more hyperedges with the same attributes (nonetheless, they are still distinct hyperedges) and may even have empty hyperedges (i.e., with no attributes at all).  $G$ is {\em clean} if $E$ has no subsumed edges. Some of our results will apply only to clean hypergraphs.

\vgap

Denote by $T$ a hyperedge tree of $G$ (the existence of $T$ is guaranteed; see Section~\ref{sec&intro&prob}). By rooting $T$ at an arbitrary leaf, we can regard $T$ as a {\em rooted} tree. Make all the links\footnote{Remember that we refrain from saying ``edges'' of $T$; see Section~\ref{sec&intro&prob}.} of $T$ point {\em downwards}, i.e., from parent to child. This way, $T$ becomes a directed acyclic graph. 

\vgap

Now that there are two views of $T$ (i.e., undirected and directed), we ought to be careful with terminology. By default, we will treat $T$ as a directed tree. Accordingly, a {\em leaf} of $T$ is a node with out-degree 0, a {\em path} is a sequence of nodes where each node has a link pointing to the next node, and a {\em subtree} rooted at a node $e$ is the directed tree induced by the nodes reachable from $e$ in $T$. Sometimes, we may revert back to the undirected view of $T$. In that case, we use the term {\em raw leaf} for a leaf in the undirected $T$ (i.e., a raw leaf can be a leaf or the root under the directed view)


\subsection{Fundamental Definitions and Properties} \label{sec&cover&basic} 

\extraspacing {\bf Summits and Disappearing Attributes.} We say that the root of $T$ is the {\em highest} node in $T$ and, in general, a node is {\em higher} (or {\em lower}) than any of its proper descendants (or ancestors). For each attribute $X \in V$, we define the {\em summit} of $X$ as the highest node (a.k.a.\ a hyperedge) that contains $X$. If node $e$ is the summit of $X$, we call $X$ a {\em disappearing} attribute in $e$. By acyclicity's connectedness requirement (Section~\ref{sec&intro&prob}), $X$ can appear only in the subtree rooted at $e$ and hence ``disappears'' as soon as we leave the subtree. 

\begin{example} 
    Let $G = (V, E)$ be the hypergraph in Example~\ref{ex&intro} whose (rooted) hypergraph tree $T$ is shown in Figure~\ref{fig&intro-ex}. The summit of \ttt{C} is node \ttt{CEJ}. Thus, \ttt{C} is a disappearing attribute of \ttt{CEJ}. Node \ttt{EHJ} is the summit of \ttt{E} and \ttt{J}. Hence, both \ttt{E} and \ttt{J} are disappearing attributes of \ttt{EHJ}. \done
\end{example}

\noindent {\bf Canonical Edge Cover.} We say that a subset $S \subseteq E$ {\em covers} an attribute $X \in V$ if $S$ has a hyperedge containing $X$. Recall that an optimal edge cover of $G$ is the smallest $S$ covering every attribute in $V$. Optimal edge covers are not unique. Some are of particular importance to us; and we will identify them as ``canonical''. Towards a procedural definition, consider the following algorithm: 

\mytab{
    \> \textsf{edge-cover} $(T)$ /* $T$ is rooted */ \\
    \> 1. \> $F_\tmp = \emptyset$ \\ 
    \> 2. \> obtain a reverse topological order $e_1, e_2, ..., e_{|E|}$ of the nodes (i.e., hyperedges) in $T$ \\ 
    \> 3. \> {\bf for} $i$ = 1 to $|E|$ {\bf do} \\ 
    \> 4. \>\> {\bf if} $e_i$ has a disappearing attribute not covered by $F_\tmp$ {\bf then} add $e_i$ to $F_\tmp$ \\
    \> 5. \> {\bf return} $F_\tmp$
}


\begin{lemma} \label{lmm&edge-cover}
    The output of {\em \textsf{edge-cover}} --- denoted as $F$ --- is an optimal edge cover of $G$, and does \uline{not} depend on the reverse topological order at Line 2. Furthermore, if $G$ is clean, $F$ includes all the raw leaves of $T$.
\end{lemma}

All the missing proofs can be found in the appendix. We refer to $F$ as the {\em canonical edge cover} (CEC) of $G$ induced by $T$. The size of $F$ is precisely the fractional edge covering number $\rho$ of $Q$. 

\begin{example} 
    Continuing on the previous example, consider the reverse topological order of $T$: \ttt{ABC}, \ttt{BD}, \ttt{BO}, \ttt{BCE}, \ttt{EFG}, \ttt{CEF}, \ttt{CEJ}, \ttt{HI}, \ttt{EHJ}, \ttt{LM}, \ttt{KL}, \ttt{HK}, \ttt{HN}. When processing \ttt{ABC}, \textsf{edge-cover} adds it to $F_\tmp$ because \ttt{ABC} has a disappearing attribute $\ttt{A}$ and yet $F_\tmp = \emptyset$. When processing \ttt{BCE}, $F_\tmp = \set{\ttt{ABC}, \ttt{BD}, \ttt{BO}}$. \ttt{BCE} has a disappearing attribute \ttt{B}, which, however, has been covered by $F_\tmp$. Thus, \ttt{B} is not added to $F_\tmp$. The final output of the algorithm is $F = \set{\ttt{ABC}$, $\ttt{BD},$ $\ttt{BO}$, $\ttt{EFG}$, $\ttt{HI},\ttt{LM}, \ttt{EHJ},  \ttt{HK}, \ttt{HN}}$, which is the CEC of $G$ induced by $T$. \done
\end{example}

\noindent {\bf Signature Paths.} Whenever $F$ includes the root of $T$, we can define a {\em signature path} --- denoted as $\sigp(f, T)$ --- for each node (i.e., hyperedge) $f \in F$. Specifically, $\sigp(f, T)$ is a set of nodes defined as follows:  
\myitems{
    \item If $f$ is the root of $T$, $\sigp(f, T) = \set{f}$. 
    \item Otherwise, let $\hat{f}$ be the lowest node in $F$ that is a proper ancestor of $f$. Then, $\sigp(f, T)$ is the set of nodes on the path from $\hat{f}$ to $f$, except $\hat{f}$. 
}

\begin{example} 
    Consider the set $F$ obtained in the previous example. If $f = \ttt{HN}$, then the signature path of $f$ is $\set{\ttt{HN}}$. If $f = \ttt{ABC}$, then $\hat{f} = \ttt{EHJ}$; and the signature path of $f$ is $\set{\ttt{ABC}, \ttt{BCE}, \ttt{CEJ}}$. \done
\end{example}

\noindent {\bf (Clean $\bm{G}$) Clustering, Anchor Leaf, and Anchor Attribute.} Consider $G = (V, E)$ now as a clean hypergraph. Let $F$ be the CEC of $G$ induced by a hyperedge tree $T$ of $G$. As $F$ contains the root and leaves of $T$ (Lemma~\ref{lmm&edge-cover}), $\set{\sigp(f, T) \mid f \in F}$ is a clustering of $E$. If $f$ is not the root of $T$, we call $\sigp(f,T)$ a {\em non-root} cluster.\footnote{If $f$ is the root of $T$, $\sigp(f,T)$ contains just $f$ itself.}


\vgap 

Let $f^\circ$ be a leaf node in $F$, and $\hat{f}$ be the lowest proper ancestor of $f^\circ$ in $F$. We call $f^\circ$ an {\em anchor leaf} of $T$ if two conditions are satisfied: 
\myitems{
    \item $\hat{f}$ has no non-leaf proper descendants in $F$. 
    \item $f^\circ$ has an attribute $A^\circ$ such that $A^\circ \notin \hat{f}$ but $A^\circ \in e$ for every node $e \in \sigp(f^\circ, T)$.
}
$A^\circ$ will be referred to as an {\em anchor attribute} of $f^\circ$. 

\begin{lemma} \label{lmm&anchor}
    If $G$ is clean, $F$ always contains an anchor leaf.
\end{lemma}

\begin{example} \label{ex&anchor}
    From the $F$ constructed earlier, we obtain the clustering $C =$ $\{\{\ttt{BO},$ $\ttt{BCE},$ $\ttt{CEJ}\},$ $\{\ttt{ABC},$ $\ttt{BCE},$ $\ttt{CEJ}\},$ $\{\ttt{BD},$ $\ttt{BCE},$ $\ttt{CEJ}\},$ $\{\ttt{EFG},$ $\ttt{CEF}$, $\ttt{CEJ}\},$ $\{\ttt{HI}\}$, $\{\ttt{EHJ}\}$, $\{\ttt{LM},$ $\ttt{KL}\},$ $\{\ttt{HK}\},$ $\{\ttt{HN}\}\}$. Other than $\set{\ttt{HN}}$, all the clusters in $C$ are non-root clusters. \ttt{ABC} is an anchor leaf of $T$ with an anchor attribute \ttt{C}. \ttt{HI} is another anchor leaf with an anchor attribute \ttt{I}. For a non-example, \ttt{BD} is not an anchor leaf because it does not have an attribute that exists in all the nodes in $\sigp(\ttt{BD}, T) = \{\ttt{BD},$ $\ttt{BCE},$ $\ttt{CEJ}\}$. Furthermore, \ttt{LM} is not an anchor leaf because \ttt{HK}, the lowest proper ancestor of \ttt{LM} in $F$, has a non-leaf proper descendant in $F$ (i.e., \ttt{EHJ}). \done
\end{example}

\subsection{(Clean $\bm{G}$) Properties on Residual Hypergraphs} 

This subsection assumes $G = (V, E)$ to be clean. Let $T$ be a hyperedge tree of $G$ and $F$ be the CEC induced by $T$. Fix an arbitrary anchor leaf $f^\circ$ of $T$ and an anchor attribute $A^\circ$ of $f^\circ$. We will analyze how the CEC changes as $G$ is simplified based on $f^\circ$ and $A^\circ$. 


\subsubsection{Simplification 1} \label{sec&cover&residual1}

The first simplification is based on removing attribute $A^\circ$ from $G$. 

\extraspacing {\bf Residual Hypergraph.} Let $G' = (V', E')$ be the {\em residual hypergraph} obtained by eliminating $A^\circ$ from $G$: $V' = V \setminus \set{A^\circ}$, and $E'$ collects a hyperedge $e' = e \setminus \set{A^\circ}$ for every $e \in E$.\footnote{If $e=\set{A^\circ}$, $E'$ collects $e' = \emptyset$.} We characterize the one-one correspondence between $E'$ and $E$ by introducing a function $\map(e) = e'$ and its inverse function $\map^{-1}(e') = e$. Let $T'$ be the hyperedge tree of $G'$ obtained by discarding $A^\circ$ from every node in $T$ (note: $G'$ is not necessarily clean).

\extraspacing {\bf Canonical Edge Cover.} Define
\myeqn{
    F' &=& 
    \left\{
    \begin{tabular}{ll}
        $F \setminus \set{f^\circ}$ & if $\map(f^\circ)$ is subsumed in $G'$ \\
        $\left\{\map(f) \mid f \in F \right\}$ & otherwise 
    \end{tabular}
    \right. 
    \label{eqn&F'}
}

\begin{example} \label{ex&residual}
    Continuing on the previous example, if we choose $f^\circ = \ttt{ABC}$ with $A^\circ =$  \ttt{C} and eliminate \ttt{C} from the tree $T$ in Figure~\ref{fig&intro-ex}, we obtain the hyperedge tree $T'$ in Figure~\ref{fig&residual}a, where the circled nodes constitute the set $F'$. Similarly, if we choose $f^\circ = \ttt{HI}$ with $A^\circ =$  \ttt{I}, then $T'$ and $F'$ are as demonstrated in Figure~\ref{fig&residual}b.
    \done
\end{example}

\begin{lemma} \label{lmm&residual1-canonical-edge-cover}
    If $G$ is clean, $F'$ is the CEC of $G'$ induced by $T'$. Furthermore, if $\map(f^\circ)$ is subsumed in $G'$, then $A^\circ$ must be an exclusive attribute in $f^\circ$.
\end{lemma}

As a corollary, if $\map(f^\circ)$ is subsumed in $G'$, then every hyperedge of $G$, except $f^\circ$, is directly retained in $G'$; furthermore, $\map(f^\circ)$ is the only subsumed edge in $G'$. The next lemma gives another property of $F'$ that holds no matter if $G$ is clean.

\begin{lemma} \label{lmm&residual1-subsumed-no-F}
    If a hyperedge $e'$ of $G'$ is subsumed, then $e' \notin F'$
\end{lemma}

\extraspacing {\bf Cleansing.} Even though $G$ is clean, the residual hypergraph $G'$ may contain subsumed hyperedges. Next, we describe a {\em cleansing} procedure which converts $G'$ into a clean hypergraph $G^* = (V', E^*)$ (note that $G^*$ has the same vertices as $G'$) and converts $T'$ into a rooted hyperedge tree $T^*$ of $G^*$. 

\vgap 

Cleansing is simple if $\map(f^\circ)$ is subsumed in $G'$. In this case, $G^*$ is the hypergraph obtained by removing $\map(f^\circ)$ from $G'$, and $T^*$ is the tree obtained by removing the leaf $\map(f^\circ)$ from $T'$. If $\map(f^\circ)$ is not subsumed, the cleansing algorithm is:


\myfigg{
    \begin{tabular}{cc}
        \includegraphics[height=40mm]{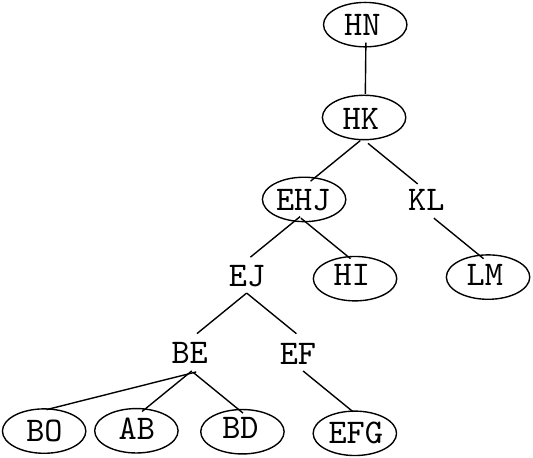} &
        \includegraphics[height=40mm]{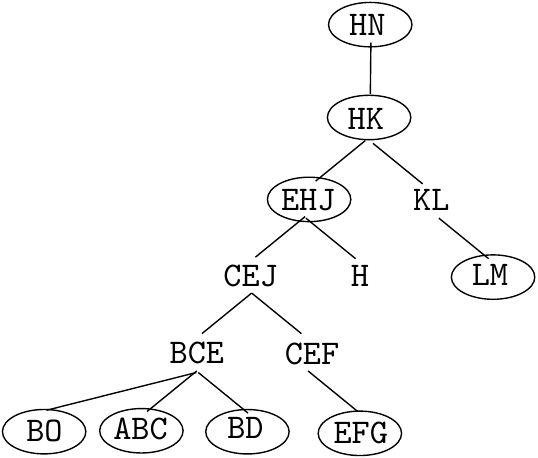} \\
        (a) $T'$ and $F'$ after removing \ttt{C} &
        (b) $T'$ and $F'$ after removing \ttt{I} \\
    \end{tabular}
}{Residual hypergraphs \label{fig&residual}}

\myfigg{
    \begin{tabular}{cc}
        \includegraphics[height=23mm]{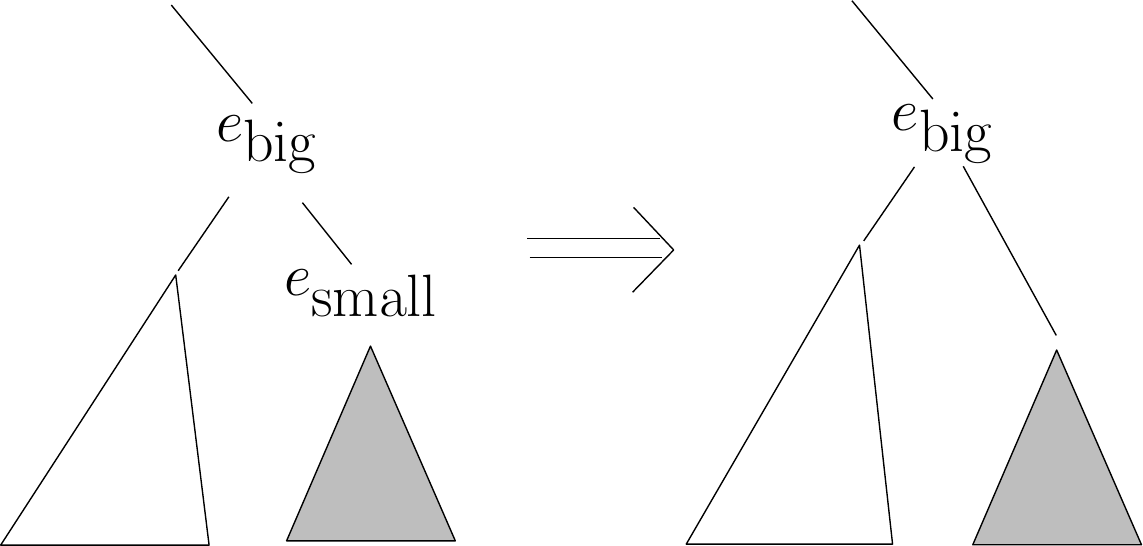}
        & \hspace{10mm}
        \includegraphics[height=23mm]{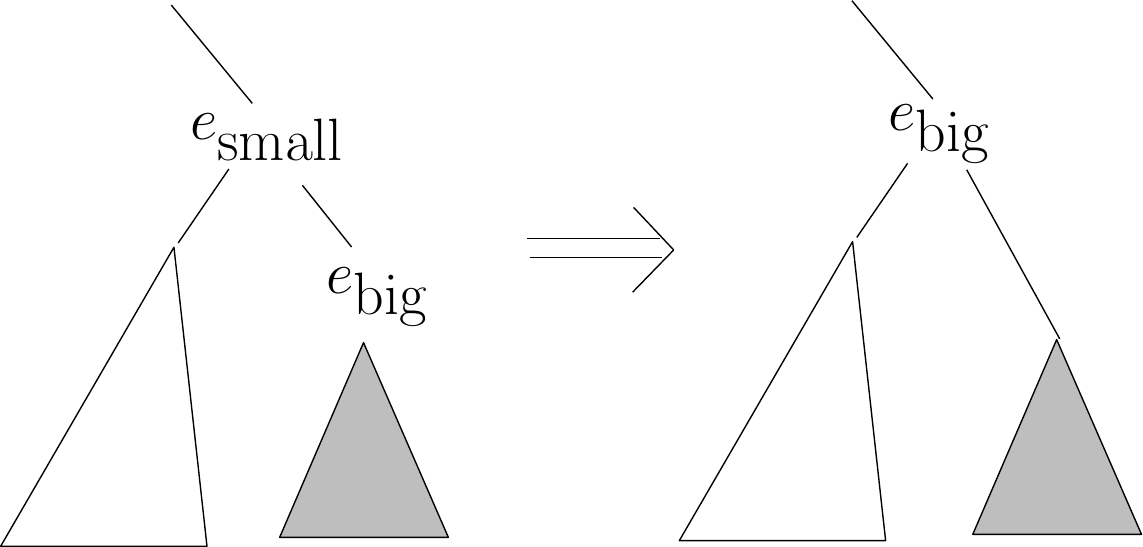} \\ 
        (a) $e_\big$ parents $e_\sml$ & \hspace{10mm} (b) $e_\sml$ parents $e_\big$
    \end{tabular}
}{Two cases of cleansing \label{fig&cleanse-2cases}}

\mytab{
    \> \textsf{cleanse} $(G', T')$ /* condition: $\map(f^\circ)$ not subsumed */ \\
    \> 1.\> $G^* = G', T^* = T'$ \\  
    \> 2. \> {\bf while} $G^*$ has hyperedges $e_\sml$ and $e_\big$ such that $e_\sml \subseteq e_\big$ and they are connected by \\ 
    \>\> a link in $T^*$ {\bf do} \\
    \> 3. \>\> remove $e_\sml$ from $G^*$ and $T^*$ \\ 
    \>\>\> /* $e_\sml \notin F'$ by Lemma~\ref{lmm&residual1-subsumed-no-F} */ \\ 
    \> 4. \>\> {\bf if} $e_\big$ was the parent of $e_\sml$ in $T^*$ {\bf then} \\
    \> 5. \>\>\> make $e_\big$ the new parent for all the child nodes of $e_\sml$; see Figure~\ref{fig&cleanse-2cases}a \\
    \>\>\> {\bf else} \\ 
    \> 6. \>\>\> make $e_\big$ the new parent for the child nodes of $e_\sml$, and\\
    \>\>\>\> make $e_\big$ a child of the (original) parent of $e_\sml$ in $T^*$; see Figure~\ref{fig&cleanse-2cases}b \\ 
    \> 7. \> {\bf return} $G^*$ and $T^*$
}

\noindent At the end of cleansing, we always set $F^* = F'$, regardless of whether $\map(f^\circ)$ is subsumed.

\begin{lemma} \label{lmm&residual1-cleansing}
    After cleansing, $F^*$ is the CEC of $G^*$ induced by $T^*$.  
\end{lemma}

\myfigg{
    \begin{tabular}{cc}
        \includegraphics[height=33mm]{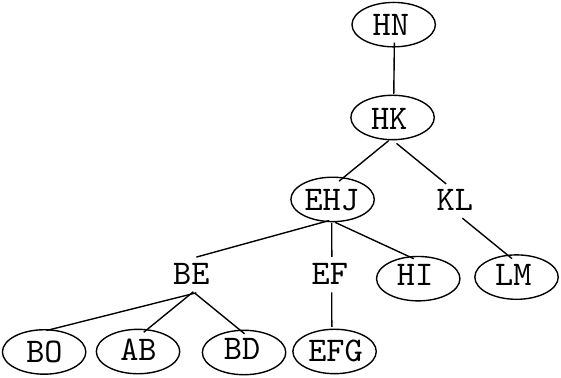} &
        \includegraphics[height=33mm]{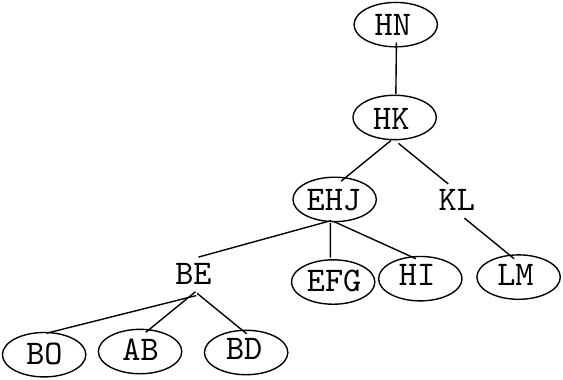} \\
        (a) After removing \ttt{EJ} &
        (b) After removing \ttt{EF} 
    \end{tabular}
}{Simplification 1 \label{fig&cleanse-ex}}

\begin{example} 
    In Example~\ref{ex&residual}, the residual hypergraph $G'$ in Figure~\ref{fig&residual}a has two subsumed hyperedges \ttt{EJ} and \ttt{EF}, each removed by an iteration of \textsf{cleanse}. Suppose that the first iteration sets $e_\sml = \ttt{EJ}$ and $e_\big = \ttt{EHJ}$ (this is a case of Figure~\ref{fig&cleanse-2cases}a). Figure~\ref{fig&cleanse-ex}a illustrates the $T^*$ after removing $\ttt{EJ}$. The next iteration sets $e_\sml = \ttt{EF}$ and $e_\big = \ttt{EFG}$ (a case of Figure~\ref{fig&cleanse-ex}b). Figure~\ref{fig&cleanse-ex}b illustrates the $T^*$ after removing $\ttt{EF}$. In both Figure \ref{fig&cleanse-ex}a and \ref{fig&cleanse-ex}b, the circled nodes constitute the CEC of $G^*$ induced by $T^*$. \done
\end{example}

\noindent {\bf Distinct Clusters Lemma.} The next property concerns the hypergraph $G^* = (V', E^*)$ after cleansing and the original hypergraph $G = (V, E)$. Recall that $T^*$ and $T$ are hyperedge trees of $G^*$ and $G$, respectively. Before proceeding, the reader should recall that every hyperedge $e^* \in E^*$ corresponds to a distinct hyperedge $e \in E$, which is the hyperedge given by $\map^{-1}(e^*)$. 

\vgap 

Consider once again the CEC $F$ of $G$, i.e., the original hypergraph, induced by $T$. As mentioned in Section~\ref{sec&cover&basic}, $C = \set{\sigp(f, T) \mid f \in F}$ is a clustering of $E$. By the same reasoning, because $F^*$ is the CEC of $G^*$ induced by $T^*$ (Lemma~\ref{lmm&residual1-cleansing}), $C^* = \set{\sigp(f^*, T^*) \mid f^* \in F^*}$ must be a clustering of $E^*$. The following lemma draws a connection between $C$ and $C^*$: 

\begin{lemma} [Distinct Clusters Lemma] \label{lmm&residual1-distinct-cluster}
    For any $1 \le k \le |F^*|$, if $\set{e^*_1, ..., e^*_k}$ is a $k$-group of $C^*$, then $\set{\map^{-1}(e^*_1), ..., \map^{-1}(e^*_k)}$ is a $k$-group of $C$. 
\end{lemma}

By definition of $k$-group, $e^*_1, ..., e^*_k$ originate from $k$ distinct clusters in $C^*$. The lemma promises $k$ different clusters in $C$ each containing a distinct hyperedge in $\set{\map^{-1}(e^*_1), ..., \map^{-1}(e^*_k)}$.

\begin{example} 
    Consider the $T^*$ (and hence $G^*$) and $F^*$ illustrated in Figure~\ref{fig&cleanse-ex}b. The clustering $C^*$ is $\{\{\ttt{AB}$, $\ttt{BE}\}$, $\{\ttt{BO}$, $\ttt{BE}\}$, $\{\ttt{BD}$, $\ttt{BE}\}$, $\{\ttt{EFG}\}$, $\{\ttt{EHJ}\}$, $\{\ttt{HI}\}$, $\{\ttt{LM}$, $\ttt{KL}\}$, $\{\ttt{HK}\}$, $\{\ttt{HN}\}\}$. Because $\{\ttt{BE}, \ttt{EFG}, \ttt{KL}\}$ is a 3-group of $C^*$, Lemma~\ref{lmm&residual1-distinct-cluster} asserts that $\{\map^{-1}(\ttt{BE}),$ $\map^{-1}(\ttt{EFG})$, $\map^{-1}(\ttt{KL})\}$ $=$ $\{\ttt{BCE}, \ttt{EFG}, \ttt{KL}\}$ must be a 3-group of the clustering $C$ in Example~\ref{ex&anchor}. \done
\end{example}

\subsubsection{Simplification 2} \label{sec&cover&residual2}

The second simplification decomposes $G$ into multiple hypergraphs based on $\sigp(f^\circ,T)$. 

\extraspacing {\bf Decomposition.} Define $Z$ to be the set of nodes $z$ in $T$ satisfying: $z$ is not in $\sigp(f^\circ,T)$ but the parent of $z$ is. For each $z \in Z$, define a rooted tree $T^*_z$ as follows: 
\myitems{
    \item The root of $T^*_z$ is the parent of $z$ in $T$. 
    \item The root of $T^*_z$ has only one child in $T^*_z$, which is $z$. 
    \item The subtree rooted at $z$ in $T^*_z$ is the same as the subtree rooted at $z$ in $T$. 
}
Separately, define $\bar{T}^*$ as the rooted tree obtained by removing from $T$ the subtree rooted at the highest node in $\sigp(f^\circ, T)$. 

\vgap

From each $T^*_z$, generate a hypergraph $G^*_z = (V^*_z, E^*_z)$. Specifically, $E^*_z$ includes all and only the nodes (each being a hyperedge) in $T^*_z$, and $V^*_z$ is the set of attributes appearing in at least one hyperedge in $E^*_z$. Likewise, from $\bar{T}^*$,  generate a hypergraph $\bar{G}^* = (\bar{V}^*, \bar{E}^*)$ where $\bar{E}^*$ includes all and only the nodes in $\bar{T}^*$, and $\bar{V}^*$ is the set of attributes appearing in at least one hyperedge in $\bar{E}^*$. 

\vgap 

Because $G$ is clean, so must be all the generated hypergraphs. Furthermore, each of them has fewer edges than $G$.\footnote{Because $f^\circ$ does not appear in any of the generated hypergraphs.} For each $z \in Z$, $T^*_z$ is a hyperedge tree of $G^*_z$; similarly, $\bar{T}^*$ is a hyperedge tree of $\bar{G}^*$. 

\begin{example} 
    In our running example, $f^\circ = \ttt{ABC}$, whose signature path is $\sigp(f^\circ, T) = \{\ttt{ABC}, \ttt{BCE}, \ttt{CEJ}\}$; see Figure~\ref{fig&decomp}a. $Z = \{\ttt{BO}, \ttt{BD}, \ttt{CEF}\}$. Figure~\ref{fig&decomp}b, \ref{fig&decomp}c, and \ref{fig&decomp}d illustrate $T^*_z$ for $z = {\ttt{CEF}}$, \ttt{BO}, and \ttt{BD}, respectively. Figure~\ref{fig&decomp}e gives $\bar{T}^*_z$. \done
\end{example}

\noindent {\bf Canonical Edge Covers.} Recall that $F$ is the CEC of $G$ induced by $T$. Next, we derive the CECs of the hypergraphs generated from the decomposition. For each $z \in Z$, define 
\myeqn{
    F^*_z &=& \set{\textrm{parent of $z$}} \cup (F \cap E^*_z).
    \label{eqn&F-star-p} 
}
Also, define
\myeqn{
    \bar{F}^* &=& F \cap \bar{E}^*.
    \label{eqn&F-star-bar} 
}

\begin{lemma} \label{lmm&residual2-edge-cover}
    For each node $z \in Z$, $F^*_z$ is the CEC of $G^*_z$ induced by $T^*_z$. Furthermore, $\bar{F}^*$ is the CEC of $\bar{G}^*$ induced by $\bar{T}^*$. 
\end{lemma}

\begin{example} 
    We have circled the nodes in $F^*_z$ in Figure~\ref{fig&decomp}b, \ref{fig&decomp}c, and \ref{fig&decomp}d for $z = \ttt{CEF}$, \ttt{BO}, and \ttt{BD}, respectively. Similarly, the circled nodes in Figure~\ref{fig&decomp}e constitute $\bar{F}^*_z$. \done
\end{example}

\noindent {\bf Distinct Clusters Lemma 2.} We close the section with a property resembling Lemma~\ref{lmm&residual1-distinct-cluster}. 

\vgap

Consider any $z \in Z$. Because $G^*_z  = (V^*_z, E^*_z)$ is clean and $F^*_z$ is the CEC of $G^*_z$ induced by $T^*_z$, $C^*_z = \set{\sigp(f^*, T^*_z) \mid f^* \in F^*_z}$ is a clustering of $E^*_z$. Similarly, regarding $\bar{G}^* = (\bar{V}^*, \bar{E}^*)$, $\bar{C}^* = \set{\sigp(f^*, \bar{T}^*) \mid f^* \in \bar{F}^*}$ is a clustering of $\bar{E}^*$. 

\myfigg{
    \begin{tabular}{ccccc}
    \includegraphics[height=40mm]{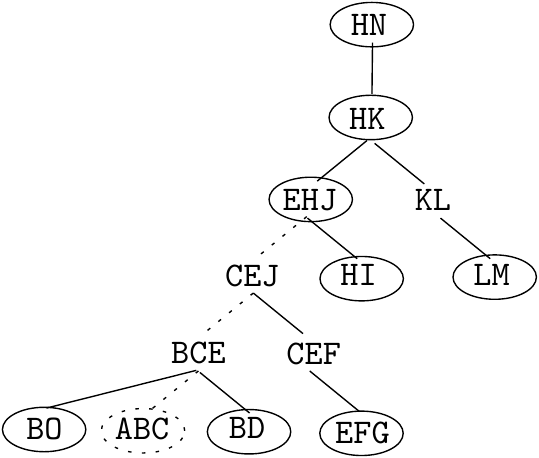} &
        \includegraphics[height=18mm]{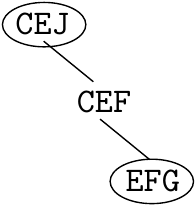} &
        \includegraphics[height=11mm]{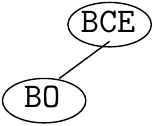} &
        \includegraphics[height=11mm]{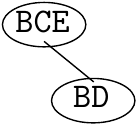} &
        \includegraphics[height=26mm]{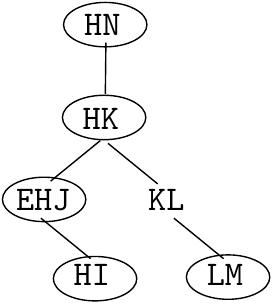} \\
        (a) The dotted path is $\sigp(\ttt{ABC},T)$ & 
        (b) $T^*_{\ttt{CEF}}$ &
        (c) $T^*_{\ttt{BO}}$ & 
        (d) $T^*_{\ttt{BD}}$ & 
        (e) $\bar{T}^*$
    \end{tabular}
}{Decomposition \label{fig&decomp}}

\vgap

Define a {\em super-$k$-group} to be a set of hyperedges $K = \set{e_1, e_2, ..., e_k}$ satisfying: 
\myitems{
    \item Each $e_i$, $i \in [k]$, is taken from a cluster of $\bar{C}^*$ or a non-root cluster\footnote{Namely, $e_i$ cannot be the root of $T^*_z$.} of $C^*_z$ for some $z \in Z$. 
    \item No two hyperedges in $K$ are taken from the same cluster. 
}
Before delving into the next lemma, the reader should recall that $\set{\sigp(f, T) \mid f \in F}$ is a clustering of $E$. 

\begin{lemma} [Distinct Clusters Lemma 2] \label{lmm&residual2-distinct-cluster}
    If $\set{e_1, e_2, ..., e_k}$ is a super-$k$-group, then $\set{e_1, e_2, ..., e_k}$ must be a $k$-group of the clustering $\set{\sigp(f, T) \mid f \in F}$. 
\end{lemma}

\begin{example} 
    In Figure~\ref{fig&decomp}, $C^*_\ttt{CEF} = \{\{\ttt{EFG}, \ttt{CEF}\}, \{\ttt{CEJ}\}\}$, $C^*_\ttt{BO} = \{\{\ttt{BO}\}, \{\ttt{BCE}\}\}$, $C^*_\ttt{BD} = \{\{\ttt{BD}\}, \{\ttt{BCE}\}\}$, $\bar{C}^* = \{\{\ttt{HI}\}, \{\ttt{EHJ}\}, \{\ttt{HK}\}, \{\ttt{HN}\}, \{\ttt{LM}, \ttt{KL}\}\}$. A super-4-group is \{\ttt{CEF}, \ttt{BO}, \ttt{BD}, \ttt{KL}\}. Lemma~\ref{lmm&residual2-distinct-cluster} assures us that \{\ttt{CEF}, \ttt{BO}, \ttt{BD}, \ttt{KL}\} must be a 4-group in the clustering $C$ given in Example~\ref{ex&anchor}. \done
\end{example}

\section{An MPC Algorithm} \label{sec&alg} 

The rest of the paper will apply the theory of CECs to solve acyclic queries in the MPC model. We will describe a variant of Hu's algorithm \cite{h21} in this section\footnote{Our algorithm follows Hu's ideas \cite{h21} but differs in certain details. For example, Hu's algorithm takes an arbitrary optimal edge cover of $G$ as the input, while we insist on working with a CEC.} and present our analysis in the next section. Denote by $Q$ the acyclic query to be answered. Let $G = (V, E)$ be the hypergraph of $Q$. We assume $G$ to be clean; otherwise, $Q$ can be converted to a clean query having the same result with load $O(m/p)$ \cite{h21}. We will also assume that $Q$ has at least two relations; otherwise, the query is trivial and requires no communication. 


\subsection{Configurations} \label{sec&alg&alg} 

Let $T$ be a hyperedge tree of $G$ and $F$ be the CEC of $G$ induced by $T$. The size of $F$ is precisely $\rho$, the fractional edge covering number of $Q$ (Section~\ref{sec&intro&prev}). As explained in Section~\ref{sec&cover&basic}, when $G$ is clean,
\myeqn{
    C &=& \set{\sigp(f, T) \mid f \in F}
    \label{eqn&C}
}
is a clustering of $E$. Let $f^\circ$ be an anchor leaf of $T$ and $A^\circ$ an anchor attribute of $f^\circ$ (Section~\ref{sec&cover&basic}); remember that $A^\circ$ appears in all the hyperedges of $\sigp(f^\circ, T)$. Define
\myeqn{
    L &=& 
    \textrm{the $Q$-induced load of $C$.}
    \label{eqn&L}
}
The reader can review Equation \eqref{eqn&clustering-load} for the definition of ``$Q$-induced load''. 

\vgap

For each hyperedge $e \in E$, as before $R(e)$ denotes the relation in $Q$ corresponding to $e$. Fix a value $x \in \dom$. Given an $e \in \sigp(f^\circ, T)$, we define the {\em $A^\circ$-frequency} of $x$ in $R(e)$ as the number of tuples $\bm{u} \in R(e)$ such that $\bm{u}(A^\circ) = x$. Further define the {\em signature-path $A^\circ$-frequency} of $x$ as the sum of its $A^\circ$-frequencies in the $R(e)$ of all $e \in \sigp(f^\circ, T)$. A value $x \in \dom$ is 
\myitems{
    \item {{\em heavy}}, if its signature-path $A^\circ$-frequency is at least $L$; 
    \item {{\em light}}, otherwise.
}
Divide $\dom$ into disjoint intervals such that the light values in each interval have a total signature-path $A^\circ$-frequency of $\Theta(L)$. We will refer to those intervals as the {\em light intervals} of $A^\circ$.
The total number of heavy values and light intervals is at most 
\myeqn{
    \sum_{e \in \sigp(f^\circ,T)} \fr{|R(e)|}{L}
    =
    O\left(\max_{e \in \sigp(f^\circ,T)} \fr{|R(e)|}{L} \right)
    =O\left(\fr{\textrm{max $(1,Q)$-product of $C$}}{L}
    \right)
    =
    O(p) \label{eqn&alg-config-num}
}
where the first equality used the fact that $\sigp(f^\circ,T)$ has $O(1)$ edges and the second equality applied the definition of max $(k,Q)$-product (see Section~\ref{sec&intro&ours}).

\vgap 

A {\em configuration} $\eta$ is either a heavy value or a light interval. Equation \eqref{eqn&alg-config-num} implies that the number of configurations is $O(p)$. For each hypergraph $e \in E$, define a relation $R(e, \eta)$ as follows:
\myitems{
    \item if $\eta$ is a heavy value, $R(e, \eta)$ includes all and only the tuples $\bm{u} \in R(e)$ satisfying $\bm{u}(A^\circ) = \eta$;
    \item if $\eta$ is a light interval, $R(e, \eta)$ includes all and only the tuples $\bm{u} \in R(e)$ where $\bm{u}(A^\circ)$ is a light value in $\eta$.
}
Note that $R(e,\eta) = R(e)$ if $A^\circ \notin e$. Let $Q_\eta$ be the query defined by $\set{R(e, \eta) \mid e \in E}$. Our objective is to compute $\join(Q_\eta)$ for all $\eta$ in parallel. The final result $\join(Q)$ is simply $\bigcup_\eta \join(Q_\eta)$. 

\vgap

The rest of the section will explain how to solve $\join(Q_\eta)$ for an arbitrary $\eta$. We allocate 
\myeqn{
    p_\eta &=&
    \Theta\left(1 + \max_{k=1}^{|F|} \fr{P_k(Q_\eta, C)}{L^k}
    \right)
    \label{eqn&p-eta}
}
machines for this purpose, where $P_k(Q_\eta, C)$ is the max $(k,Q_\eta)$-product of $C$.

\subsection{Solving $\bm{Q_\eta}$ When $\bm{\eta}$ is a Heavy Value} \label{sec&alg-heavy} 

Define the residual hypergraph $G' = (V', E')$ after removing $A^\circ$, and also functions $\map(.)$ and $\map^{-1}(.)$ as in Section~\ref{sec&cover&residual1}. We compute $\join(Q_\eta)$ in five steps. 

\vgap 

\uline{\em Step 1.} Send the tuples of $R(e, \eta)$, for all $e \in E$, to the $p_\eta$ allocated machines such that each machine receives $\Theta(\fr{1}{p_\eta} \sum_{e \in E} |R(e, \eta)|)$ tuples. 

\vgap 
    
    \uline{\em Step 2.} For each $e \in E$, convert $R(e, \eta)$ to $R^*(e', \eta)$ where $e' = \map(e) = e \setminus \set{A^\circ}$. Specifically, $R^*(e', \eta)$ is a copy of $R(e, \eta)$ but with $A^\circ$ discarded, or formally, $R^*(e', \eta) = \set{\bm{u}[e'] \mid \textrm{tuple } \bm{u} \in R(e, \eta)}$. No communication occurs as each machine simply discards $A^\circ$ from every tuple $\bm{u} \in R(e, \eta)$ in the local storage. 
    
\vgap 
    
    \uline{\em Step 3.} Cleanse $G'$ into $G^* = (V', E^*)$. As explained in Section~\ref{sec&cover&residual1}, this may or may not require calling algorithm \textsf{cleanse}. If called,  \textsf{cleanse} identifies in each iteration two hyperedges $e_\sml$ and $e_\big$ in the current $G^*$ and removes $e_\sml$. Accordingly, we perform a {\em semi-join} between $R^*(e_\sml, \eta)$ and $R^*(e_\big, \eta)$, which removes every tuple $\bm{u}$ from $R^*(e_\big, \eta)$ with the property that $\bm{u}[e_\sml]$ is absent from $R^*(e_\sml, \eta)$. $R^*(e_\sml, \eta)$ is discarded after the semi-join. 
    
\vgap 
    
    \uline{\em Step 4.} Let $Q^*_\eta$ be the query defined by the relation set $\set{R^*(e^*, \eta) \mid e^* \in E^*}$. Compute $\join(Q^*_\eta)$ using $p_\eta$ machines recursively. Note that the number of participating attributes has decreased by 1 for the recursion. 
    
\vgap 
    
    \uline{\em Step 5.} We output $\join(Q_\eta)$ by augmenting each tuple $\bm{u} \in \join(Q^*_\eta)$ with $\bm{u}(A^\circ) = \eta$. No communication is needed.

\subsection{Solving $\bm{Q_\eta}$ When $\bm{\eta}$ is a Light Interval} \label{sec&alg-light} 


Define $Z$, $G^*_z = (V^*_z, E^*_z)$ (for each $z \in Z$), $C^*_z$, $\bar{G}^* = (\bar{V}^*, \bar{E}^*)$, and $\bar{C}^*$ all in the way described in Section~\ref{sec&cover&residual2}. We compute $\join(Q_\eta)$ in four steps.

\vgap 

\uline{\em Step 1.} Same as Step 1 of the algorithm in Section~\ref{sec&alg-heavy}. 

\vgap

\uline{\em Step 2.} For each $e \in \sigp(f^\circ, T)$, broadcast $R(e,\eta)$ to all $p_\eta$ machines. By definition of light interval, the size of $R(e,\eta)$ is at most $L$. 

\vgap

\uline{\em Step 3.} For each $z \in Z$, define a query $Q^*_{\eta, z} = \set{R(e, \eta) \mid e \in E^*_z}$. Similarly, for $\bar{G}^*$, define a query $\bar{Q}^*_\eta  = \set{R(e, \eta) \mid e \in \bar{E}^*}$. Next, we compute the cartesian product of $\join(\bar{Q}^*_\eta )$ and the $\join(Q^*_{\eta, z})$ of all the $z \in Z$ --- namely $\left(\times_{z \in Z} \join(Q^*_{\eta, z}) \right) \times \join(\bar{Q}^*_\eta )$ --- using $p_\eta$ machines. Towards that purpose, define for each $z \in Z$
\myeqn{
    p_{\eta,z} &=&
    \Theta\left(1 + \max_{k=1}^{|F^*_z|} \fr{P_k(Q^*_{\eta, z}, C^*_z)}{L^k}
    \right)
    \label{eqn&p-eta-z}
}
where $P_k(Q^*_{\eta, z}, C^*_z)$ is the max $(k, Q^*_{\eta, z})$-product of the clustering $C^*_z$. Similarly, define 
\myeqn{
    \bar{p}_\eta &=&
    \Theta\left(1 + \max_{k=1}^{|\bar{F}^*|} \fr{P_k(\bar{Q}^*_\eta, \bar{C}^*)}{L^k}
    \right)
    \label{eqn&p-eta-bar}
}
where $P_k(\bar{Q}^*_\eta, \bar{C}^*)$ is the max $(k, \bar{Q}^*_\eta)$-product of the clustering $\bar{C}^*$. We will prove later that each $Q^*_{\eta, z}$ can be answered with load $O(L)$ using $p_{\eta,z}$ machines, and $\bar{Q}^*_\eta $ can be answered with load $O(L)$ using $\bar{p}_\eta$ machines. Therefore, applying the cartesian product algorithm given in Lemma 6 of \cite{ks17} (see also Lemma 4 of \cite{kst20}), we can compute $\left(\times_{z \in Z} \join(Q^*_{\eta, z}) \right) \times \join(\bar{Q}^*_\eta )$ with load $O(L)$ using $\bar{p}_\eta \cdot \prod_{z \in Z} p_{\eta,z}$ machines. As proved later, we can adjust the constants in \eqref{eqn&p-eta-z} and \eqref{eqn&p-eta-bar} to make sure $\bar{p}_\eta \cdot \prod_{z \in Z} p_{\eta,z} \le p_\eta$, where $p_\eta$ is given in \eqref{eqn&p-eta}.

\vgap

\uline{\em Step 4.} We combine the cartesian product $\left(\times_{z \in Z} \join(Q^*_{\eta, z}) \right) \times \join(\bar{Q}^*_\eta )$ with the tuples broadcast in Step 2 to derive $\join(Q_\eta)$ with no more communication. Specifically, for each tuple $\bm{u}$ in the cartesian product, the machine where $\bm{u}$ resides outputs $\set{\bm{u}} \bowtie \left(\bowtie_{e \in \sigp(f^\circ, T)} R(e, \eta)\right)$. It is rudimentary to verify that all the tuples of $\join(Q_\eta)$ will be produced this way.

\section{Analysis of the Algorithm} \label{sec&analysis}

This section will establish: 

\begin{theorem} \label{thm&main}
    Consider any join query $Q$ defined in Section~\ref{sec&intro&prob} whose hypergraph is $G$. The algorithm of Section~\ref{sec&alg} answers $Q$ with load $O(L)$, where $L$ (given in \eqref{eqn&L}) is the $Q$-induced load of the clustering obtained from a canonical edge cover of $G$. 
\end{theorem}

We will prove the theorem by induction on the number of participating attributes (i.e., $|V|$) and the number of participating relations (i.e., $|Q|$). If $|Q| = 1$, the theorem trivially holds. If $|V| = 1$, $Q$ has only one relation (because $Q$ is clean) and the theorem also holds. Next, assuming that the theorem holds on any query with {\em either} strictly less participating attributes {\em or} strictly less participating relations than $Q$, we will prove the theorem's correctness on $Q$.

\vgap 

Our analysis will answer three questions. First, why do we have enough machines to handle all configurations in parallel? In particular, we must show that $\sum_\eta p_\eta \le p$, where $p_\eta$ is given in \eqref{eqn&p-eta}. Second, why does each step in Section~\ref{sec&alg-heavy} and \ref{sec&alg-light} entail a load of $O(L)$? Third, why do we have $\bar{p}_\eta \cdot \prod_{z \in Z} p_{\eta,z} \le p_\eta$ in Step 3 of Section~\ref{sec&alg-light}? Settling these questions will complete the proof of Theorem~\ref{thm&main}.

\vgap 

All the notations in this section follow those in Section~\ref{sec&alg}.

\subsection{Total Number of Machines for All Configurations} \label{sec&analysis&overall-machine}


It suffices to prove $\sum_\eta p_\eta = O(p)$ because adjusting the hidden constants then ensures $\sum_\eta p_\eta \le p$. For every $k \in [|F|]$, we will show 
\myeqn{
    \fr{1}{L^k}\sum_\eta P_k(Q_\eta,C) &=& O(p)
    \label{eqn&machine-target-eqn}
}
which will yield 
\myeqn{
    \sum_\eta p_\eta &=& 
    \sum_\eta O\left(1 + \max_{k=1}^{|F|} \fr{P_k(Q_\eta,C)}{L^k}\right) \nn \\
    &=& \sum_\eta O\left(1 + \sum_{k=1}^{|F|} \fr{P_k(Q_\eta,C)}{L^k} \right) 
    = O(p) + \sum_{k=1}^{|F|} O\left(\sum_\eta \fr{P_k(Q_\eta,C)}{L^k} \right) \nn \\ 
    &=& O(p) \nn
}
where the second equality used  $|F| = O(1)$ and the third equality used $\sum_\eta 1 = O(p)$.\footnote{ $\sum_\eta 1$ is the number of configurations which is $O(p)$ as shown in \eqref{eqn&alg-config-num}.} 

\vgap

Henceforth, fix the value of $k$. For any $\eta$, the hypergraph of $Q_\eta$ is always $G$ (i.e., the hypergraph of $Q$). Consider an arbitrary $k$-group $K$ of the clustering $C$ (given in Equation~\ref{eqn&C}). The $Q_\eta$-product of $K$ is $\prod_{e \in K} |R(e,\eta)|$.\footnote{For the definition of ``a $k$-group's $Q$-product'', review Section~\ref{sec&intro&ours}.} For any $K$, we will prove   
\myeqn{
    \fr{1}{L^k} \sum_\eta \prod_{e \in K} |R(e,\eta)| &=& O(p). 
    \label{eqn&machine-target-eqn2}
}
As $C$ has $O(1)$ $k$-groups $K$, the above yields 
\myeqn{
    \sum_\eta \fr{P_k(Q_\eta,C)}{L^k} &=& \sum_\eta \fr{1}{L^k} \max_{K} \prod_{e \in K} |R(e,\eta)| \nn \\ 
    &=& O\left(\sum_\eta \fr{1}{L^k}  \sum_K \prod_{e \in K} |R(e,\eta)|\right) 
    = O\left(\sum_K \fr{1}{L^k}  \sum_\eta \prod_{e \in K} |R(e,\eta)|\right) \nn \\ 
    &=& \sum_{K} O(p) = O(p) \nn 
}
as claimed in \eqref{eqn&machine-target-eqn}.


\vgap 

Let us first consider the case where $K \cap \sigp(f^\circ, T) \ne \emptyset$, namely, $K$ has a hyperedge $e_0$ picked from the cluster $\sigp(f^\circ, T)$. We have: 
\myeqn{
    \sum_\eta \prod_{e \in K} |R(e,\eta)| 
    &=&
    \sum_\eta \Big( |R(e_0,\eta)| \cdot \prod_{e \in K \setminus \set{e_0}} |R(e,\eta)|
    \Big) \label{eqn&machine-target-eqn3-help1}
}
For each $e \in K \setminus \set{e_0}$, obviously $|R(e,\eta)| \le |R(e)|$. Regarding $e_0$, because $A^\circ$ must be an attribute of $e_0$, the relations $R(e_0, \eta)$ of all the configurations $\eta$ form a {\em partition} of $R(e_0)$.\footnote{The $R(e_0, \eta)$ of all the $\eta$ are mutually disjoint and their union equals $R(e_0)$.} Hence:
\myeqn{
    \eqref{eqn&machine-target-eqn3-help1}
    &\le&
    \Big(\prod_{e \in K \setminus \set{e_0}} |R(e)| \Big) \Big(\sum_\eta |R(e_0,\eta)|\Big)
    =
    \Big(\prod_{e \in K \setminus \set{e_0}} |R(e)| \Big) \cdot |R(e_0)|  
    =
    \prod_{e \in K} |R(e)| 
    \nn \\
    &\le&
    \textrm{max $(k,Q)$-product of $C$}. \nn 
}
Therefore, the left hand side of \eqref{eqn&machine-target-eqn2} is bounded by $\fr{\textrm{max $(k,Q)$-product of $C$}}{L^k}$, which is at most $p$ (by definition of $L$).  

\vgap 

Next, we consider $K \cap \sigp(f^\circ, T) = \emptyset$. In this case, we must have $k = |K| \le |F| - 1$, because the hyperedges in $K$ need to come from distinct clusters of $C$, and $C$ has $|F|$ clusters (one of them is $\sigp(f^\circ, T)$, which now must be excluded). Applying the trivial fact $|R(e,\eta)| \le |R(e)|$ (for any $e$) and the fact that $\sum_\eta 1$ is bounded by \eqref{eqn&alg-config-num}, we have 
\myeqn{
    \fr{1}{L^k} \sum_\eta \prod_{e \in K} |R(e,\eta)|
    &\le& 
    \fr{1}{L^k} \sum_\eta \prod_{e \in K} |R(e)| 
    = 
    O\left(
    \fr{1}{L^k} \prod_{e \in K} |R(e)| \cdot \max_{e \in \sigp(f^\circ, T)} \fr{|R(e)|}{L}
    \right) \nn \\ 
    &=&
    O\left(
    \fr{\textrm{max $(k+1,Q)$-product of $C$}}{L^{k+1}}
    \right) \nn 
}
which is at most $p$. This completes the proof of $\sum_\eta p_\eta = O(p)$.

\subsection{Heavy $\bm{Q_\eta}$} \label{sec&analysis&heavy}

This subsection will prove that the algorithm in Section~\ref{sec&alg-heavy} has load $O(L)$. Step 2 and 5 demand no communication. The loads of Step 1 and 3 can all be bounded\footnote{Step 3 performs $O(1)$ semi joins, each of which can be performed by sorting. For sorting in the MPC model, see Section 2.2.1 of \cite{hyt19}. The stated bound for Step 1 and 3 requires the assumption $p \le m^{1-\eps}$.} by $O(\fr{1}{p_\eta} \sum_{e \in E} |R(e, \eta)|) = O(\fr{1}{p_\eta} \max_{e \in E} |R(e, \eta)|) = O(P_1(Q_\eta,C) / p_\eta) = O(L)$. 

\vgap 

To analyze Step 4, let $T^*$ be the hyperedge tree of $G^*$ (produced by cleansing) and $F^*$ be the CEC of $G^*$. By definition, the $Q^*_\eta$-induced load of the clustering $C^* = \set{\sigp(f^*, T^*) \mid f^* \in F^*}$ is
\myeqn{
    L^*_\eta &=& \max_{k=1}^{|F^*|} \left(\fr{P_k(Q^*_\eta,C^*)}{p_\eta}\right)^{1/k} \label{eqn&L*_eta} 
}
where $P_k(Q^*_\eta,C^*)$ is the max $(k,Q^*_\eta)$-product of $C^*$.    By our inductive assumption (that Theorem~\ref{thm&main} holds on $Q^*_\eta$), Step 4 incurs load $O(L^*_\eta)$. We will prove $P_k(Q^*_\eta,C^*) \le P_k(Q_\eta,C)$ for every $k$ which, together with \eqref{eqn&p-eta} and \eqref{eqn&L*_eta}, will tell us $L^*_\eta = O(L)$.

\vgap 

Before proceeding, the reader should recall that, for any hyperedge $e^*$ of $G^*$, $\map^{-1}(e^*)$ gives a hyperedge in $G$. We must have $|R^*(e^*, \eta)| \le |R(\map^{-1}(e^*), \eta)|$. To see why, note that this is true when $|R^*(e^*, \eta)|$ is created in Step 2, whereas $R^*(e^*, \eta)$ can only shrink in Steps 3-5.  

\vgap 

To prove $P_k(Q^*_\eta,C^*) \le P_k(Q_\eta,C)$, consider any $k$-group $K^*$ of $C^*$. By Lemma~\ref{lmm&residual1-distinct-cluster}, $K = \set{\map^{-1}(e^*) \mid e^* \in K^*}$ must be a $k$-group of $C$. Since $|R^*(e^*,\eta)| \le |R(\map^{-1}(e^*),\eta)|$ for any $e^* \in K^*$, we have $\prod_{e^* \in K^*} |R^*(e^*, \eta)| \le \prod_{e \in K} |R(e, \eta)| \le P_k(Q_\eta,C)$. Therefore:
\myeqn{
    P_k(Q^*_\eta,C^*) = \max_{K^*} \prod_{e^* \in K^*} |R^*(e^*, \eta)| \le P_k(Q_\eta,C).
    \nn
} 

\subsection{Light $\bm{Q_\eta}$} \label{sec&analysis&light}

This subsection will concentrate on the algorithm of Section~\ref{sec&alg-light}.

\extraspacing {\bf Load.} Step 1 incurs load $O(L)$ (same analysis as in Section~\ref{sec&alg-heavy}). Step 2 also requires a load of $O(L)$ because every broadcast relation has a size of at most $L$. Step 4 needs no communication. 

\vgap

To analyze Step 3, let us first consider $\bar{Q}^*_\eta$. The $\bar{Q}^*_\eta$-induced load of the clustering $\bar{C}^*$ is 
\myeqn{
    \bar{L}^*_\eta &=& 
    \max_{k=1}^{|\bar{C}^*|} \left(\fr{P_k(\bar{Q}^*_\eta,\bar{C}^*)}{\bar{p}_\eta} \right)^{1/k} \nn
}
where $P_k(\bar{Q}^*_\eta,\bar{C}^*)$ as the max $(k,\bar{Q}^*_\eta)$-product of $\bar{C}^*$. By our inductive assumption (that Theorem~\ref{thm&main} holds on $\bar{Q}^*_\eta$), answering $\bar{Q}^*_\eta $ with $\bar{p}_\eta$ machines requires load $O(\bar{L}^*_\eta)$, which is $O(L)$ given the $\bar{p}_\eta$ in \eqref{eqn&p-eta-bar}. A similar argument shows that answering each $Q^*_{\eta, z}$ with $p_{\eta,z}$ machines --- with $p_{\eta,z}$ given in \eqref{eqn&p-eta-z} --- incurs a load of $O(L)$. Thus, the cartesian product at Step 3 can be computed with load $O(L)$.  

\extraspacing {\bf Number of machines in Step 3.}  Next, we will prove that $\bar{p}_\eta \cdot \prod_{z \in Z} p_{\eta,z} \le p_\eta$ always holds in Step 3. It suffices to show $\bar{p}_\eta \cdot \prod_{z \in Z} p_{\eta,z} = O(p_\eta)$ which, as we will see, relies on Lemma~\ref{lmm&residual2-distinct-cluster} and the fact that $|R(e,\eta)| \le L$ for every node $e \in \sigp(f^\circ, T)$. 

\vgap

Consider an arbitrary $z \in Z$. The root of $T^*_z$ --- denoted as $e_\root$ --- must belong to $\sigp(f^\circ, T)$. Recall that a $k$-group $K$ of $C^*_z$ takes a hyperedge from a distinct cluster in $C^*_z$. Call $K$ a {\em non-root $k$-group} if $e_\root \notin K$, or a {\em root $k$-group}, otherwise. Define 
\myeqn{
    P_k(Q^*_{\eta, z},C^*_z)
    &=& \text{max $(k,Q^*_{\eta, z})$-product of $C^*_z$} \nn \\ 
    P^\non_k(Q^*_{\eta,z},C^*_z)
    &=& \text{max $(k,Q^*_{\eta, z})$-product of all the {\em non-root} $k$-groups of $C^*_z$}. \nn 
}
As a special case, define $P^\non_0(Q^*_{\eta,z},C^*_z) = 1$.
For any $k$, we observe 
\myeqn{
    P_k(Q^*_{\eta,z},C^*_z) 
    &\le& 
    \max\{P^\non_k(Q^*_{\eta,z},C^*_z), L \cdot P^\non_{k-1}(Q^*_{\eta,z},C^*_z)\}. 
    \label{eqn&analysis-step3-help-1}
}
To prove the inequality, fix $K$ to the $k$-group with the largest $Q^*_{\eta,z}$-product ($= P_k(Q^*_{\eta,z},C^*_z)$). If $K$ is a non-root $k$-group, \eqref{eqn&analysis-step3-help-1} obviously holds. Consider, instead, that $K$ is a root $k$-group. Since $e_\root \in \sigp(f^\circ, T)$, we know $|R(e_\root, \eta)| \le L$ and hence $\prod_{e \in K} |R(e, \eta)| \le L \cdot \prod_{e \in K \setminus \set{e_\root}} |R(e, \eta)|$. As $K \setminus \set{e_\root}$ is a non-root $(k-1)$-group, $P_k(Q^*_{\eta,z},C^*_z) \le L \cdot P^\non_{k-1}(Q^*_{\eta,z},C^*_z)$ holds. 

\vgap 

Equipped with \eqref{eqn&analysis-step3-help-1}, we can now derive from \eqref{eqn&p-eta-z}:
\myeqn{
    p_{\eta,z} 
    &=&
    O\left(1 + \max_{k=1}^{|F^*_z|} \fr{\max\{P^\non_k(Q^*_{\eta,z},C^*_z), L \cdot P^\non_{k-1}(Q^*_{\eta,z},C^*_z)\}}{L^k}
    \right) \nn \\
    &=&
    O\left(1 + \max_{k=1}^{|F^*_z|-1} \fr{P^\non_k(Q^*_{\eta,z},C^*_z)}{L^k}
    \right) 
    \label{eqn&analysis-step3-help-2}
}
where the second equality used the fact that, when $k = |F^*_z|$, a $k$-group must be a root $k$-group. 


We are now ready to prove $\bar{p}_\eta \cdot \prod_{z \in Z} p_{\eta,z} = O(p_\eta)$. For each $z \in Z$, define integer $k_z$ and a set $K_z$ of hyperedges as follows: 
\myitems{
    \item If \eqref{eqn&analysis-step3-help-2} $=\Theta(P^\non_k(Q^*_{\eta,z},C^*_z)/L^k)$ for some $k \in [1, |F^*_z|-1]$, 
    set $k_z = k$ and $K_z$ to the non-root $k$-group whose $Q^*_{\eta,z}$-product equals $P^\non_k(Q^*_{\eta,z},C^*_z)$. 
    \item Otherwise (we must have $p_{\eta,z} = \Theta(1)$), set $k_z = 0$ and $K_z = \emptyset$; furthermore, define the $Q^*_{\eta,z}$-product of $K_z$ to be 1.  
}

Similarly, regarding $\bar{p}_\eta$ in~\eqref{eqn&p-eta-bar}, define integer $\bar{k}$ and a set $\bar{K}$ of hyperedges as follows: 
\myitems{
    \item If \eqref{eqn&p-eta-bar} $=\Theta(P_k(\bar{Q}^*_\eta,\bar{C}^*)/L^k)$ for some $k \in [1, |\bar{F}^*|]$, set $\bar{k} = k$ and $\bar{K}$ to the $k$-group of the clustering $\bar{C}^*$ whose $\bar{Q}^*_\eta$-product equals $P_k(\bar{Q}^*_\eta,\bar{C}^*)$. 
    \item Otherwise, set $\bar{k} = 0$ and $\bar{K} = \emptyset$; furthermore, define the $\bar{Q}^*_\eta$-product of $\bar{K}$ to be 1.  
}

Define $K_\super = \bar{K} \cup \left(\bigcup_{z \in Z} K_z \right)$. If $K_\super = \emptyset$, then $p_{\eta,z} = \Theta(1)$ for all $z \in Z$ and $\bar{p}_\eta = \Theta(1)$, which leads to 
\myeqn{
    \bar{p}_\eta \cdot \prod_{z \in Z} p_{\eta,z} = O(1) = O(p_\eta). 
    \nn
}
If $K_\super \ne \emptyset$, $K_\super$ is a super-$|K_\super|$-group\footnote{For the definition of super-$k$-group, review Section~\ref{sec&cover&residual2}.}. By
Lemma~\ref{lmm&residual2-distinct-cluster}, $K_\super$ is a $|K_\super|$-group of $T$. We thus have: 
\myeqn{
    \bar{p}_\eta \cdot \prod_{z \in Z} p_{\eta,z} 
    &=&
    \fr{\textrm{$\bar{Q}^*_\eta$-product of $\bar{K}$}}{L^{\bar{k}}} 
    \prod_{z \in Z} \fr{\textrm{$Q^*_{\eta,z}$-product of $K_z$}}{L^{k_z}} \nn \\
    &=&
    \fr{\prod_{e \in K_\super} |R(e)|}{L^{|K_\super|}}
    \le
    \fr{\textrm{max $(|K_\super|,Q_\eta)$-product of $C$}}{L^{|K_\super|}}
    =
    O(p_\eta). \nn  
}
This completes the whole proof of Theorem~\ref{thm&main}.

\bibliographystyle{plain}
\bibliography{ref}

\appendix 

\section*{Appendix} 

\section{Proof of Lemma~\ref{lmm&edge-cover}} \label{app&lmm-edge-cover-alg} 

We first show that $F$ is an edge cover of $G$. Each attribute $X \in V$ is a disappearing attribute of some hyperedge $e \in E$. When $e$ is processed at Line 4 of \textsf{edge-cover}, either $X$ is already covered or $e$ itself will be added to $F_\tmp$ (which will then cover $X$).

\vgap 

Next, we argue that $F$ is an {\em optimal} edge cover (i.e., having the smallest size). Let $F'$ be an arbitrary optimal edge cover of $G$. We will establish a one-one mapping between $F$ and $F'$, which implies the optimality of $F$. 

\vgap 

Fix an arbitrary hyperedge $e \in F$. If $e$ also belongs to $F'$, we map $e$ to its copy in $F'$. Consider the opposite case where $e \notin F'$. The fact $e \in F$ indicates that when 
\textsf{edge-cover} processes $e$, $e$ must contain a disappearing attribute $X$ that has not been covered by $F_\tmp$. Let $e' \in F'$ be an arbitrary hyperedge containing $X$; we map $e$ to $e'$. As explained in Section~\ref{sec&cover&basic}, $e'$ must be a proper descendant of $e$ in $T$. 

\vgap 

We argue that no two $e$ and $\hat{e}$ in $F$ can be mapped to the same hyperedge $e' \in F$. If this happens, $e'$ is a descendant of both $e$ and $\hat{e}$. Assume, without loss of generality, that $e$ is a proper descendant of $\hat{e}$. Since $\hat{e}$ is mapped to $e'$, there is an attribute $Y$ such that
\myitems{
    \item $Y$ is a disappearing attribute in $\hat{e}$ not covered by $F_\tmp$ when \textsf{edge-cover} adds $\hat{e}$ to $F_\tmp$; 
    \item $Y \in e'$. 
}
Because $e$ is on the path from $\hat{e}$ to $e'$ in $T$, connectedness of acyclicity guarantees $Y \in e$. On the other hand, $e \in F$ and $e$ is processed before $\hat{e}$ (reverse topological order). Thus, when $\hat{e}$ is processed, $e \in F_\tmp$ and hence $Y$ must be covered by $F_\tmp$, giving a contradiction. 

\vgap 

We now proceed to show that $F$ does not depend on the reverse topological order at Line 2. Recall that, when processing a node $e$, \textsf{edge-cover} adds it to $F_\tmp$ if and only if $F_\tmp$ does not cover a disappearing attribute $X$ of $e$. All the nodes containing $X$ must appear in the subtree of $T$ rooted at $e$ and thus must be processed before $e$. Hence, whether $e \in F$ is determined by which of those nodes are selected into $F_\tmp$. The observation gives rise to an inductive argument. First, if $e$ is a leaf, $e$ enters $F_\tmp$ if and only if it has a disappearing attribute (which must be exclusive), independent of the reverse topological order used. For a non-leaf node $e$, inductively, once we have decided whether $e' \in F_\tmp$ for every proper descendent $e'$ of $e$, whether $e \in F_\tmp$ has also been decided. We thus conclude that the reverse topological order has no influence on the output.

\vgap 

It remains to show that when $G$ is clean, $F$ must include all the raw leaf nodes $e$ of $T$. If $e$ is not the root of $T$, it must have an attribute $X$ absent from the parent node of $e$ (otherwise, $e$ is subsumed by its parent and $G$ is not clean). Similarly, if $e$ is the root of $T$, it must  have an attribute $X$ absent from its child (there is only one child because $e$ is a raw leaf). In both cases, the attribute $X$ is exclusive at $e$ and will force \textsf{edge-cover} to add $e$ to $F_\tmp$. 

\section{Proof of Lemma~\ref{lmm&anchor}} \label{app&lmm-anchor} 

Identify an arbitrary non-leaf node $\hat{f} \in F$ such that no other non-leaf node in $F$ is lower than $\hat{f}$. The existence of $\hat{f}$ is guaranteed because $F$ includes the root of $T$. Consider any child node $e$ of $\hat{f}$ in $T$. Since $G$ is clean, $e$ must have an attribute $A^\circ$ that does not appear in $\hat{f}$. Let $f^\circ$ be any node in $F$ that contains $A^\circ$. By the connectedness requirement of acyclicity, $f^\circ$ must be in the subtree of $T$ rooted at $e$ and, therefore, must be a leaf.   

\vgap 

We argue that $f^\circ$ is an anchor leaf. The signature path of $f^\circ$ includes all the nodes on the path from $e$ to $f^\circ$. Because $A^\circ \in e$ and $A^\circ \in f^\circ$, $A^\circ$ must appear in all the nodes on the path (connectedness requirement) and is thus an anchor attribute of $f^\circ$.

\section{Proof of Lemma~\ref{lmm&residual1-canonical-edge-cover}} \label{app&lmm-residual-canonical-edge-cover} 

\subsection{$\bm{\map(f^\circ)}$ Subsumed in $\bm{G'}$}

Let $\hat{e}$ be the parent of $f^\circ$ in $T$. If $\map(f^\circ) = f^\circ \setminus \{A^\circ\}$ is subsumed in $G'$, then $\map(f^\circ)$ must be a subset of $\map(\hat{e})$, which indicates $A^\circ \notin \hat{e}$ (otherwise, $f^\circ \subseteq \hat{e}$ and $G$ is not clean). Because $A^\circ$ needs to appear in all the nodes of $\sigp(f^\circ, T)$, $A^\circ \notin \hat{e}$ indicates that $\sigp(f^\circ, T)$ has only a single node $f^\circ$. It thus follows that $\hat{e} \in F$ and $A^\circ$ is an exclusive attribute in $f^\circ$. Hence, the removal of $A^\circ$ does not affect any hyperedge except $f^\circ$. 


\vgap

Next, we show that $F' = F \setminus \{f^\circ\}$ is the CEC of $G'$ induced by $T'$. It suffices to prove that $F'$ is the output of \textsf{edge-cover}$(T')$ on some reverse topological order of $T'$. For this purpose, consider $\sigma_0$ as an arbitrary reverse topological order of $T$ where $\hat{e}$ succeeds $f^\circ$. Let $\sigma_1$ be the sequence obtained by removing $f^\circ$ from $\sigma_0$; $\sigma_1$ must be a reverse topological order of $T'$. Let $e_\bfr$ be the node preceding $f^\circ$ in $\sigma_0$ (and hence preceding $\hat{e}$ in $\sigma_1$); define $e_\bfr$ to be a dummy node if $f^\circ$ is the first in $\sigma_0$. 

\vgap 

Let us compare the execution of \textsf{edge-cover}$(T)$ on $\sigma_0$ to that of \textsf{edge-cover}$(T')$ on $\sigma_1$. The two executions are identical till the moment when $e_\bfr$ has been processed. By the fact that \textsf{edge-cover}$(T)$ adds $\hat{e}$ to $F_\tmp$ (we have proved earlier $\hat{e} \in F$), $\hat{e}$ has a disappearing attribute not covered by $F_\tmp$ when $\hat{e}$ is processed. Hence, when $\hat{e}$ is processed by \textsf{edge-cover}$(T')$, it must also have a disappearing attribute not covered by $F_\tmp$ and thus is added to $F_\tmp$. The rest execution of \textsf{edge-cover}$(T)$ is the same as that of \textsf{edge-cover}$(T')$ because every non-exclusive attribute of $f^\circ$ is in $\hat{e}$. Therefore, the output of \textsf{edge-cover}$(T')$ is the same as that of \textsf{edge-cover}$(T)$, except that the former does not include $f^\circ$.

\subsection{$\bm{\map(f^\circ)}$ Not Subsumed in $\bm{G'}$} 

Let $\sigma_0 = (e_1, e_2, ..., e_{|E|})$ be an arbitrary reverse topological order of $T$. Define $e_i' = \map(e_i) = e_i \setminus \set{A^\circ}$ for $i \in [|E|]$. The sequence $\sigma_1 = (e_1', e_2', ..., e_{|E|}')$ is a reverse topological order of $T'$. We will compare the execution of \textsf{edge-cover}$(T)$ on $\sigma_0$ to that of \textsf{edge-cover}$(T')$ on $\sigma_1$. Define $F_0(e_i)$ (resp., $F_1(e_i')$) as the content of $F_\tmp$ after \textsf{edge-cover}$(T)$ (resp., \textsf{edge-cover}$(T')$) has processed $e_i$ (resp., $e_i'$). 

\boxminipg{0.9\linewidth}{
    {\bf Claim 1:} For any leaf $e$ of $T$, {\em \textsf{edge-cover}}$(T')$ must add $e' = \map(e)$ to $F_\tmp$.
}

    Let us prove the claim. Because $e$ is a leaf of $T$ and $G$ is clean, $e$ must have an exclusive attribute $X$. If \textsf{edge-cover} does not add $e'$ to $F_\tmp$, $e'$ has no exclusive attributes in $T'$. This implies $X = A^\circ$, which further implies $f^\circ = e$ (otherwise, $A^\circ$ appears in two distinct nodes and cannot be exclusive). However, in that case, $e'$ must contain an exclusive attribute in $T'$ ($e' = \map(f^\circ)$ is not subsumed in $G'$). We thus have reached a contradiction.

\vgap

To establish Lemma~\ref{lmm&residual1-canonical-edge-cover}, it suffices to prove: 
\boxminipg{0.9\linewidth}{
    {\bf Claim 2:} For each $i$, $e_i \in F_0(e_i)$ if and only if $e_i' \in F_1(e_i')$.  
} 
We prove the claim by induction on $i$. Because $e_1$ is a leaf of $T$,  Lemma~\ref{lmm&edge-cover} and Claim 1 guarantee $e_1 \in F_0(e_1)$ and $e_1' \in F_1(e_1')$, respectively. Thus, Claim 2 holds for $i = 1$. 

\vgap

Next, we prove the correctness on $i > 1$, assuming that it holds on $e_{i-1}$ and $e'_{i-1}$. The inductive assumption implies that $F_0(e_{i-1})$ covers an attribute $X \ne A^\circ$ if and only if $F_1(e_{i-1}')$ covers $X$. If $e_i \notin F_0(e_i)$, every disappearing attribute of $e_i$ must be covered by $F_0(e_{i-1})$. Hence, $F_1(e_{i-1}')$ must cover all the disappearing attributes of $e_i'$ and thus $e_i' \notin F_1(e_i')$. 

\vgap 

The rest of the proof assumes $e_i \in F_0(e_i)$, i.e., $e_i$ has a disappearing attribute $X$ not covered by $F_0(e_{i-1})$. If $X \ne A^\circ$, $X$ is a disappearing attribute in $e_i'$ not covered by $F_1(e_{i-1}')$ and thus $e_i' \in F_1(e_i')$. It remains to discuss the scenario $X = A^\circ$. As $A^\circ$ is disappearing at $e_i$, $A^\circ$ cannot exist in the parent of $e_i$. On the other hand, because $A^\circ \in f^\circ$, acyclicity's connectedness requirement forces $f^\circ$ to be a descendant of $e_i$. We can safely conclude that $f^\circ = e_i$; otherwise, the leaf $f^\circ$ is processed before $e_i$ and must exist in $F_0(e_{i-1})$ (Lemma~\ref{lmm&edge-cover}), contradicting the fact that $A^\circ$ is not covered by $F_0(e_{i-1})$. Then, $e'_i \in F_1(e_i')$ follows from Claim 1.

\section{Proof of Lemma~\ref{lmm&residual1-subsumed-no-F}} 

    Define $e = \map^{-1}(e')$. Because $e'$ is subsumed, we know that $e$ must contain $A^\circ$ (otherwise, $e$ is subsumed and $G$ is not clean). In other words, $e = e' \cup \set{A^\circ}$. Furthermore, if $e = f^\circ$, then $\map(f^\circ) = \map(e) = e'$ is subsumed in $G'$, in which case we must have $e' \notin F'$ (by the way we define $F'$ in \eqref{eqn&F'}).    
    Next, we assume $e \ne f^\circ$. To complete the proof, it suffices to show that $e \notin F$, where $F$ is the CEC of $G$ induced by $T$. 
    
    \vgap
    
    Let $\hat{f}$ be the lowest proper ancestor of $f^\circ$ in $F$ (here, ``ancestor'' is defined with respect to $T$). By definition of $A^\circ$, $A^\circ \notin \hat{f}$. Because $A^\circ \in f^\circ$ and $A^\circ \in e$, $e$ must be a proper descendant of $\hat{f}$ in $T$. Assume, for contradiction purposes, that $e \in F$. As $\hat{f}$ (by definition of $f^\circ$) cannot have any non-leaf proper descendant in $F$, $e$ must be a leaf of $T$.
    
    \vgap
    
    Because $A^\circ$ appears in two distinct nodes $e$ and $f^\circ$, acyclicity's connected requirement demands that $A^\circ$ should also exist in the parent $\hat{e}$ of $e$. Because $G$ is clean, we know that $e$ must have at least one attribute $X$ that does not appear in $\hat{e}$ and thus must be exclusive. It follows that $X \ne A^\circ$. However, in that case $e' = e \setminus \set{A^\circ}$ contains $X$ and thus cannot be subsumed in $G'$ ($X$ remains exclusive in $G'$), giving a contradiction.

\section{Proof of Lemma~\ref{lmm&residual1-cleansing}} \label{app&lmm-residual-cleansing} 

We discuss only the scenario where $\map(f^\circ)$ is not subsumed in $G'$ (the opposite case is easy and omitted). Our proof will establish a stronger claim: 

\boxminipg{0.9\linewidth}{
    {\bf Claim:} $F^* = F'$ is the CEC of $G^*$ induced by $T^*$ every time Line 2 of \textsf{cleanse} is executed. 
} 

$G^* = G'$ and $T^* = T'$ at Line 1. $F^* = F'$ is the CEC of $G^*$ induced by $T^*$ at this moment (Lemma~\ref{lmm&residual1-canonical-edge-cover}). Hence, the claim holds on the first execution of Line 2. 

\vgap 

Inductively, assuming that the claim holds currently, we will show that it still does after \textsf{cleanse} deletes the next $e_\sml$ from $G^*$. Let $G^*_0$ and $T^*_0$ (resp., $G^*_1$ and $T^*_1$) be the $G^*$ and $T^*$ before (resp., after) the deletion of $e_\sml$, respectively. The fact $e_\sml$ being subsumed in $G^*$ suggests $e_\sml$ being subsumed in $G'$. By Lemma~\ref{lmm&residual1-subsumed-no-F}, $e_\sml \notin F' = F^*$. 

\extraspacing {\bf Case 1: $\bm{e_\big}$ parents $\bm{e_\sml}$.} Let $\sigma_0$ be a reverse topological order of $T^*_0$ where $e_\big$ succeeds $e_\sml$. As $F^*$ is the CEC of $G^*_0$ induced by $T^*_0$, \textsf{edge-cover}$(T^*_0)$ produces $F^*$ if executed on $\sigma_0$ (Lemma~\ref{lmm&edge-cover}). 

\vgap 

Let $\sigma_1$ be a copy of $\sigma_0$ but with $e_\sml$ removed; $\sigma_1$ is a reverse topological order of $T^*_1$. Every node in $T^*_1$ retains the same disappearing attributes as in $T^*_0$ (see Figure~\ref{fig&cleanse-2cases}a). For every node $e \ne e_\sml$, running \textsf{edge-cover}$(T^*_1)$ on $\sigma_1$ has the same effect as running \textsf{edge-cover}$(T^*_0)$ on $\sigma_0$. Therefore, \textsf{edge-cover}$(T^*_1)$ also outputs $F^*$.

\extraspacing {\bf Case 2: $\bm{e_\sml}$ parents $\bm{e_\big}$.} Let $\sigma_0$ be a reverse topological order of $T^*_0$ where $e_\sml$ succeeds $e_\big$. Let $\sigma_1$ be a copy of $\sigma_0$ but with $e_\sml$ removed; $\sigma_1$ is a reverse topological order of $T^*_1$. We will argue that running \textsf{edge-cover}$(T^*_1)$ on $\sigma_1$ returns $F^*$. 

\vgap

The reader should note several preliminary facts about disappearing attributes. If an attribute has $e_\sml$ as the summit in $T^*_0$, the attribute's summit in $T^*_1$ becomes $e_\big$ (see Figure~\ref{fig&cleanse-2cases}b). If an attribute has $e \ne e_\sml$ as the summit in $T^*_0$, its summit in $T^*_1$ is still $e$. Hence, every node in $T^*_1$ except $e_\big$ retains the same disappearing attributes as in $T^*_0$, whereas the disappearing attributes of $e_\big$ in $T^*_1$ contain those of $e_\big$ and $e_\sml$ in $T^*_0$.

\vgap 

For each node $e$ in $\sigma_0$ (resp.\ $\sigma_1$), denote by $F_0(e)$ (resp.\ $F_1(e)$) the content of $F_\tmp$ after \textsf{edge-cover}$(T^*_0)$ (resp.\ \textsf{edge-cover}$(T^*_1)$) has processed $e$. Let $e_\bfr$ be the node before $e_\big$ in $\sigma_0$.\footnote{In the special case where $e_\big$ is the first in $\sigma_0$, define $e_\bfr$ as a dummy node with $F_0(e_\bfr) = F_1(e_\bfr) = \emptyset$.} It is easy to see that \textsf{edge-cover}$(T^*_0)$ and \textsf{edge-cover}$(T^*_1)$ behave the same way until finishing with $e_\bfr$, which gives $F_0(e_\bfr) = F_1(e_\bfr)$. It must hold that $e_\sml \notin F_0(e_\sml)$.\footnote{Otherwise, $e_\sml \in F^*$, contradicting Lemma~\ref{lmm&residual1-subsumed-no-F}.} Two possibilities apply to $e_\big$: 

\myenums{
    \item $e_\big \in F_0(e_\big)$. Hence, $e_\big$ has a disappearing attribute in $T^*_0$ not covered by $F_0(e_\bfr)$. This means that $e_\big$ also has a disappearing attribute in $T^*_1$ not covered by $F_1(e_\bfr) = F_0(e_\bfr)$. It follows that $e_\big \in F_1(e_\big)$, meaning $F_1(e_\big) = F_0(e_\big) = F_0(e_\sml)$. 
    
    \vspace{3mm}
    
    \item $e_\big \notin F_0(e_\big)$. All the disappearing attributes of $e_\big$ and $e_\sml$ in $T^*_0$ are covered by $F_0(e_\bfr)$. Hence, the disappearing attributes of $e_\big$ in $T^*_1$ are covered by $F_1(e_\bfr) = F_0(e_\bfr)$. Therefore, $e_\big \notin F_1(e_\big)$, meaning $F_0(e_\sml) = F_0(e_\bfr) = F_1(e_\bfr) = F_1(e_\big)$. 
}

We now conclude that $F_1(e_\big) = F_0(e_\sml)$ always holds. Every remaining node in $\sigma_0$ and $\sigma_1$ has the same disappearing attributes in $T^*_0$ and $T^*_1$. The rest execution of \textsf{edge-cover}$(T^*_0)$ is identical to that of \textsf{edge-cover}$(T^*_1)$.  

\section{Proof of Lemma~\ref{lmm&residual1-distinct-cluster}} \label{app&lmm-residual1-clustering-subset}

We will discuss only the scenario where $\map(f^\circ)$ is not subsumed (the opposite scenario is easy and omitted). 

\vgap

Departing from acyclic queries, let us consider a more general problem on a rooted tree $\T$ where (i) every node is colored black or white, and (ii) the root and all the leaves are black. Denote by $B$ the set of black nodes. Each black node $b \in B$ is associated with a {\em signature path}: 
\myitems{
    \item If $b$ is the root of $\T$, its signature path contains just $b$ itself.
    \item Otherwise, let $\hat{b}$ be the lowest ancestor of $b$ among all the nodes in $B$; the signature path of $b$ is the set of nodes on the path from $\hat{b}$ to $b$, except $\hat{b}$. 
}

We define four types of contractions: 

\myitems{
    \item Type 1: We are given two white nodes $v_1$ and $v_2$ such that $v_1$ parents $v_2$. The contraction removes $v_2$ from $\T$ and makes $v_1$ the new parent for all the child nodes of $v_2$. See Figure~\ref{fig&contract}a. 
    
    \item Type 2: We are given two white nodes $v_1$ and $v_2$ such that $v_1$ parents $v_2$. The contraction removes $v_1$ from $\T$, makes $v_2$ the new parent for all the child nodes of $v_1$, and makes $v_2$ a child of the original parent of $v_1$. See Figure~\ref{fig&contract}b. 
    
    \item Type 3: Same as Type 1, except that $v_1$ is black and $v_2$ is white. See Figure~\ref{fig&contract}c.
    
    \item Type 4: Same as Type 2, except that $v_1$ is white and $v_2$ is black. See Figure~\ref{fig&contract}d. 
}

\myfigg{
    \begin{tabular}{cc}
        \includegraphics[height=25mm]{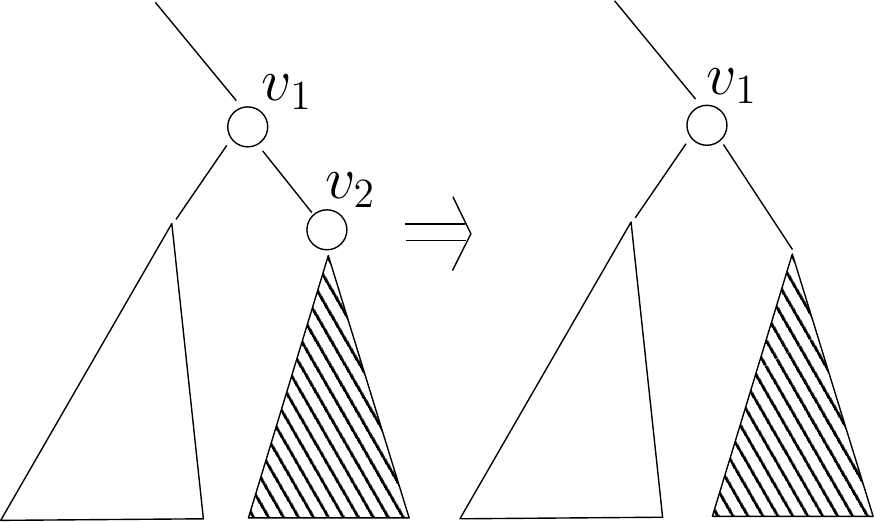}
        & \hspace{3mm}
        \includegraphics[height=25mm]{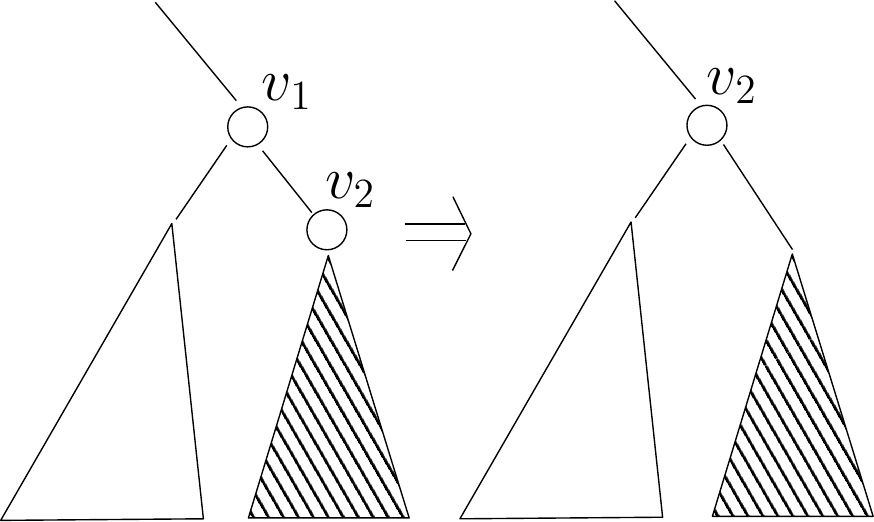} \\
        (a) Type 1 & 
        \hspace{3mm} 
        (b) Type 2 \\ 
        \includegraphics[height=25mm]{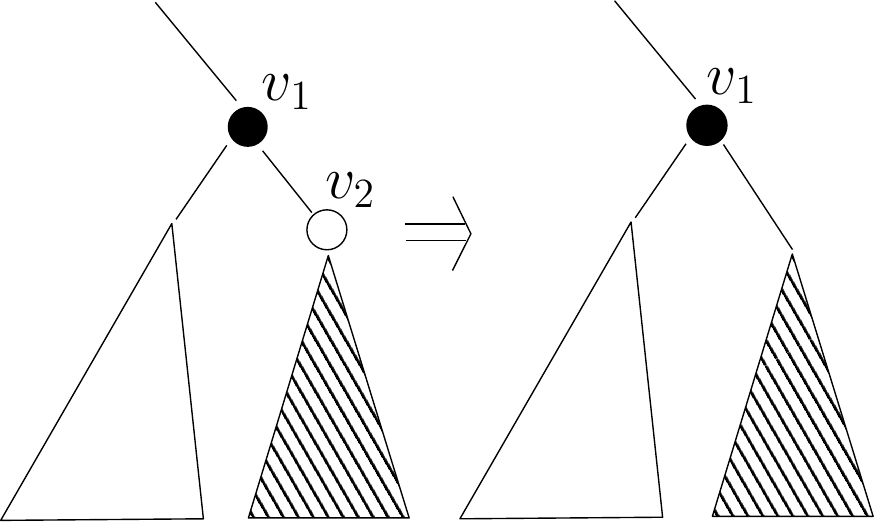} &
        \hspace{3mm} \includegraphics[height=25mm]{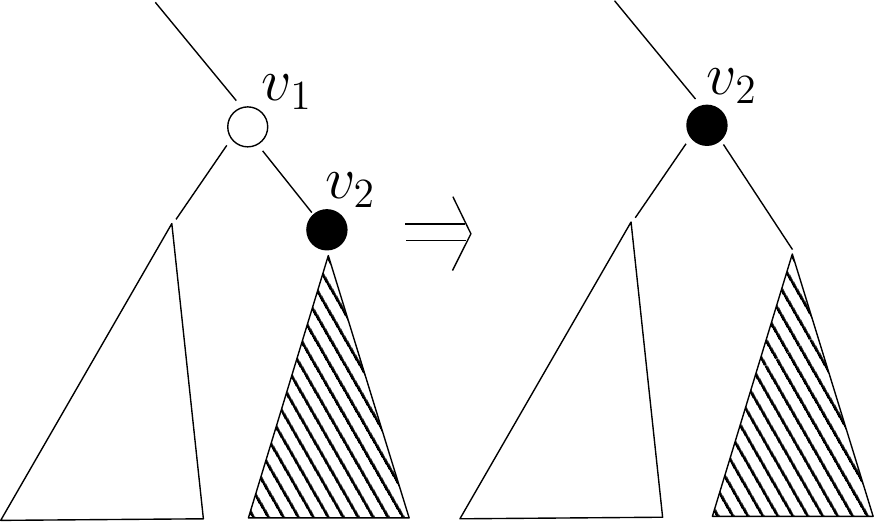} \\ 
        (c) Type 3 & 
        \hspace{3mm} 
        (d) Type 4 
    \end{tabular}
}{Four types of contraction \label{fig&contract}}


The facts below are evident: 
\myitems{
    \item The number of black nodes remains the same after a contraction. 
    \item After a contraction, each signature path either remains the same or shrinks. 
}

We now draw correspondence between a contraction and a hyperedge deletion in  \textsf{cleanse}. $\T$ corresponds to the current hyperedge tree $T^*$ in \textsf{cleanse}. The set $B$ of black nodes equals $F^* = F'$ for the entire execution of \textsf{cleanse}. The set $\set{v_1, v_2}$ corresponds to $\set{e_\sml, e_\big}$. As shown in Lemma~\ref{lmm&residual1-subsumed-no-F}, $e_\sml$ cannot exist in $F^*$ and thus cannot correspond to a black node. If we denote by $C$ (resp., $C^*$) the set of signature paths at the beginning (resp., end) of \textsf{cleanse}, each signature path in $C^*$ is obtained by continuously shrinking a distinct signature path in $C$. This implies Lemma~\ref{lmm&residual1-distinct-cluster}, noticing that $C =  \set{\sigp(f, T) \mid f \in F}$ and $C^* = \set{\sigp(f^*, T^*) \mid f^* \in F^*}$. 

\section{Proof of Lemma~\ref{lmm&residual2-edge-cover}} \label{app&lmm-residual2-edge-cover} 

We will first prove that, for any $z \in Z$, $F^*_z$ is the CEC of $G^*_z$ induced by $T^*_z$. Let $\hat{z}$ be the parent of $z$. Recall that $F$ is the CEC of $G$ induced by $T$. Consider a reverse topological order $\sigma_z$ of $T$ satisfying the following condition: a prefix of $\sigma_z$ is a permutation of the nodes in the subtree of $T$ rooted at $z$. In other words, in $\sigma_z$, every node in the aforementioned subtree must rank before every node outside the subtree. Define $\sigma^*_z$ to be the sequence obtained by deleting from $\sigma_z$ all the nodes $e$ such that $e \neq \hat{z}$ and $e$ is outside the subtree of $T$ rooted at $z$. It is clear that $\sigma^*_z$ is a reverse topological order of $T^*_z$.

\vgap

Let us compare the execution of \textsf{edge-cover}$(T)$ on $\sigma$ to that of \textsf{edge-cover}$(T^*_z)$ on $\sigma^*_z$. They are exactly the same until $z$ has been processed. Hence, every node in the $F_\tmp$ of \textsf{edge-cover}$(T)$ at this moment must have been added to $F_\tmp$ by \textsf{edge-cover}$(T^*_z)$. This means that all the nodes in $F^*_z$, except $\hat{z}$, must appear in the final $F_\tmp$ output by \textsf{edge-cover}$(T^*_z)$. Finally, the final $F_\tmp$ must also contain $\hat{z}$ as well due to Lemma~\ref{lmm&edge-cover} (notice that $\hat{z}$ is a raw leaf of $T^*_z$). This shows that $F^*_z$ is the CEC of $G^*_z$ induced by $T^*_z$.

\vgap 

Next, we prove that $\bar{F}^*$ is the CEC of $\bar{G}^*$ induced by $\bar{T}^*$. Let $\bar{e}$ be the highest node in $\sigp(f^\circ,T)$. Consider a reverse topological order $\bar{\sigma}$ of $T$ satisfying the following condition: a prefix of $\bar{\sigma}$ is a permutation of the nodes in the subtree of $T$ rooted at $\bar{e}$. Define $\bar{\sigma}^*$ to be the sequence obtained by deleting that prefix from $\bar{\sigma}$. It is clear that $\bar{\sigma}^*$ is a reverse topological order of $\bar{T}^*$. Define $\hat{\bar{e}}$ to be the parent of $\bar{e}$ in $T$. Note that $\hat{\bar{e}}$ must belong to $F$ due to the definitions of $\bar{e}$ and $\sigp(f^\circ, T)$

\vgap 

We will compare the execution of \textsf{edge-cover}$(T)$ on $\sigma$ to that of \textsf{edge-cover}$(\bar{T}^*)$ on $\bar{\sigma}^*$. For each $e$ in $\sigma$, define $F_0(e)$ as the content of $F_\tmp$ after \textsf{edge-cover}$(T)$ has finished processing $e$. Similarly, for each $e$ in $\bar{\sigma}^*$, define $F_1(e)$ as the content of $F_\tmp$ after \textsf{edge-cover}$(\bar{T}^*)$ has finished processing $e$. Divide $\sigma$ into three segments: (i) $\sigma_1$, which includes the prefix of $\sigma$ ending at (and including) $\bar{e}$, (ii) $\sigma_2$, which starts right after $\sigma_1$ and ends at (and includes) $\hat{\bar{e}}$, and (iii) $\sigma_3$, which is the rest of $\sigma$. Note that $\bar{\sigma}^*$ is the concatenation of $\sigma_2$ and $\sigma_3$. 

\boxminipg{0.9\linewidth}{
    {\bf Claim 1:} For any $e$ in $\sigma_2$, $e \in F_0(e)$ if and only if $e \in F_1(e)$. 
}

We prove the claim by induction. As the base case, consider $e$ as the first element in $\sigma_2$. In $\bar{T}^*$, $e$ must be a leaf and, by Lemma~\ref{lmm&edge-cover}, must be in $F_1(e)$. In $T$, $e$ is either a leaf or $\hat{\bar{e}}$. In the former case, Lemma~\ref{lmm&edge-cover} assures us $e \in F_0(e)$. In the latter case, $e$ is also in $F_0(e)$ because $\hat{\bar{e}} \in F$.  

\vgap 

Next, we prove the claim on every other node $e$ in $\sigma_2$, assuming the claim's correctness on the node $e_\bfr$ preceding $e$ in $\sigma_2$. This inductive assumption implies $F_1(e_\bfr) \subseteq F_0(e_\bfr)$. If $e \in F_0(e)$, then $e$ has a disappearing attribute $X$ not covered by $F_0(e_\bfr)$. As $F_1(e_\bfr) \subseteq F_0(e_\bfr)$, $F_1(e_\bfr)$ does not cover $X$, either. Hence, \textsf{edge-cover}$(\bar{T}^*)$ adds $e$ to $F_\tmp$, namely, $e \in F_1(e)$. 

\vgap 

Let us now focus on the case where $e \in F_1(e)$. If $e = \hat{\bar{e}}$, the fact $\hat{\bar{e}} \in F$ indicates $e \in F_0(e)$. Next, we consider $e \ne \hat{\bar{e}}$, meaning that $e$ is a proper descendant of $\hat{\bar{e}}$. The fact $e \in F_1(e)$ suggests that $e$ has a disappearing attribute $X$ not covered by $F_1(e_\bfr)$. If $e \notin F_0(e)$, $F_0(e_\bfr)$ must have a node $e'$ containing $X$. Node $e'$ must come from $\sigma_1$ (the inductive assumption prohibits $e'$ from appearing in $\sigma_2$) and hence must be a descendant of $\bar{e}$. By acyclicity's connectedness requirement, $X$ appearing in both $e$ and $e'$ means that $X$ must belong to $\hat{\bar{e}}$. But this contradicts $X$ disappearing at $e$. We thus conclude that $e \in F_0(e)$. 

\boxminipg{0.9\linewidth}{
    {\bf Claim 2:} For any $e$ in $\sigma_3$, $e \in F_0(e)$ if and only if $e \in F_1(e)$. 
}

Claim 1 assures us that $F_1(\hat{\bar{e}}) \subseteq F_0(\hat{\bar{e}})$. Note also that $\hat{\bar{e}}$ belongs to $F_0(\hat{\bar{e}})$ (as explained before, $\hat{\bar{e}} \in F$) and hence also to $F_1(\hat{\bar{e}})$ (Claim 1). Any node $e' \in F_0(\hat{\bar{e}}) \setminus F_1(\hat{\bar{e}})$ must appear in the subtree rooted at $\hat{\bar{e}}$ in $T$, whereas any node $e$ in $\sigma_3$ must be outside that subtree. By acyclicity's connectedness requirement, if $e'$ contains an attribute $X$ in $e$, then $X \in \hat{\bar{e}}$ for sure. This means that $F_1(\hat{\bar{e}})$ covers a disappearing attribute of $e$ if and only if $F_0(\hat{\bar{e}})$ does so. Therefore, \textsf{edge-cover}$(\bar{T}^*)$ processes each node of $\sigma_3$ in the same way as \textsf{edge-cover}$(T)$. This proves the correctness of Claim 2. 

\vgap 

By putting Claim 1 and 2 together, we conclude that \textsf{edge-cover}$(\bar{T}^*)$ returns  all and only the attributes in $\sigma_2 \cup \sigma_3$ output by \textsf{edge-cover}$(T)$. Therefore, the output of \textsf{edge-cover}$(\bar{T}^*)$ is $F \cap \bar{E}^* = \bar{F}^*$.

\section{Proof of Lemma~\ref{lmm&residual2-distinct-cluster}} \label{app&lmm-residual2-clustering-subset}

For any $f^* \in F^*_z$ and any $z \in Z$ that is not the root of $T^*_z$, it holds that $\sigp(f^*, T^*_z) \subseteq \sigp(f^*, T)$. Similarly, for any $f^* \in \bar{F}^*$, it holds that $\sigp(f^*, \bar{T}^*) \subseteq \sigp(f^*, T)$. To prove the lemma, it suffices to show that, given a super-$k$-group $K = \set{e_1, ..., e_k}$, we can always assign each $e_i$, $i \in [k]$, to a distinct cluster in $\set{\sigp(f, T) \mid f \in F}$. This is easy: if $e_i$ is picked from $\sigp(f^*, T^*_z)$ for some $z \in F$ and $f^* \in F^*_z$, assign $e_i$ to $\sigp(f^*, T)$; if $e_i$ is picked from $\sigp(f^*, \bar{T}^*)$ for some $f^* \in \bar{F}^*$, assign $e_i$ to $\sigp(f^*, T)$. 

\end{sloppy}
\end{document}